\documentclass[twocolumn]{aastex63}

\usepackage{natbib}
\usepackage{color}
\usepackage{mathtools}
\usepackage{hyperref} 
\graphicspath{{./figures/}} 

\def    \apjl  		{\rm {ApJL}}

\def    \apjl  		{\rm {ApJL}}

\def	\cm		{\,{\rm {cm}}}
\def	\K		{\,{\rm {K}}}
\def	\g		{\,{\rm {g}}}
\def	\mum	{\,{\mu \rm{m}}}

\def \bea {\begin{eqnarray}}
\def \ena {\end{eqnarray}}                  

\def    \ba     {\bf  a}

\def	\ba	{{\bf a}}

\def	\B	{{\rm B}}

\def	\C	{{\rm C}}

\def	\cm	{\,{\rm cm}}
\def	\d	{{\rm d}}

\def	\D	{{\rm D}}

\def	\erg	{\,{\rm erg}}

\def	\g	{\,{\rm g}}
\def	\gas	{\,{\rm gas}}

\def	\G	{{\rm G}}

\def	\H	{{\rm H}}

\def	\N	{{\rm N}}

\def	\s	{\,{\rm s}}

\def	\V	{{\rm V}}

\def	\rad	{\,{\rm rad}}

\def	\xhat		{\hat{\bf x}}
\def	\yhat		{\hat{\bf y}}

\def	\ba			{{\bf a}}

\def    \gas     	{{\rm gas}}

\begin{document}
\shorttitle{Ro-thermal desorption mechanism}
\shortauthors{Hoang and Tung}
\title{Chemistry on Rotating Grain Surface: Ro-Thermal Desorption of Molecules from Ice Mantles}
\author{Thiem Hoang}
\affil{Korea Astronomy and Space Science Institute, Daejeon 34055, Republic of Korea; \href{mailto:thiemhoang@kasi.re.kr}{thiemhoang@kasi.re.kr}}
\affil{University of Science and Technology, Korea, (UST), 217 Gajeong-ro Yuseong-gu, Daejeon 34113, Republic of Korea}
\author{Ngo-Duy Tung}
\affil{University of Science and Technology of Hanoi, VAST, 18 Hoang Quoc Viet, Hanoi, Vietnam}

\begin{abstract}
It is widely believed that water and complex organic molecules (COMs) first form in the ice mantle of dust grains and are subsequently returned into the gas due to grain heating by intense radiation of protostars. Previous research on the desorption of molecules from the ice mantle assumed that grains are at rest which is contrary to the fact that grains are suprathermally rotating as a result of their interaction with an anisotropic radiation or gas flow. {To clearly understand how molecules are released in to the gas phase, the effect of grain suprathermal rotation on surface chemistry must be quantified}. In this paper, we study the effect of suprathermal rotation of dust grains spun-up by radiative torques on the desorption of molecules from icy grain mantles around protostars. We show that centrifugal potential energy due to grain rotation reduces the potential barrier of molecules and significantly enhances their desorption rate. We term this mechanism {\it rotational-thermal} or {\it ro-thermal} desorption. We apply the ro-thermal mechanism for studying the desorption of molecules from icy grains which are simultaneously heated to high temperatures and spun-up to suprathermal rotation by an intense radiation of protostars. We find that ro-thermal desorption is much more efficient than thermal desorption for molecules with high binding energy such as water and COMs. Our results have important implications for understanding the origin of COMs detected in star-forming regions and call for attention to the effect of suprathermal rotation of icy grains to use molecules as a tracer of physical conditions of star-forming regions. 
\end{abstract}

\section{Introduction}\label{sec:intro}
To date, more than 200 different molecules, including water and complex organic molecules (COMs, having more than six atoms), were detected in the interstellar medium (see e.g., \citealt{Caselli:2012fq} for a review). It is thought that COMs form in the ice mantle of dust grains during the warming up phase induced by protostars (see \citealt{Herbst:2009go} for a review). However, the question of how such molecules are returned into the gas remains unclear (\citealt{2018IAUS..332....3V}). 

Several desorption mechanisms have been proposed to explain the desorption of molecules from the icy grain mantle, including thermal and non-thermal mechanisms (see \cite{vanDishoeck:2014cu} for a review). Thermal sublimation (\citealt{1972ApJ...174..321W}; \citealt{1985A&A...144..147L}; \citealt{1992ApJS...82..167H}) is the most popular mechanism to explain the detection of water and COMs in hot cores/corinos around massive/low-mass protostars because in these regions icy grains can be heated to high temperatures above 100 K (\citealt{Herbst:2009go}; \citealt{Caselli:2012fq}). Nevertheless, COMs are frequently detected in lukewarm envelopes around protostars where the temperature is below their sublimation threshold ($T\sim 50-100\K$; \citealt{2013ApJ...771...95O}; \citealt{Fayolle:2015cu}; \citealt{vanDishoeck:2013en}; \citealt{Oberg:2016el}). This casts doubt on the proxy of thermal sublimation. Moreover, non-thermal desorption mechanisms include desorption induced by cosmic rays such as whole grain heating or impulsive heating (CRs; \citealt{1985A&A...144..147L}), desorption by single UV photons (photodesorption; \citealt{2009ApJ...693.1209O}). The UV photodesorption is a promising mechanism to desorb molecules in cold dark clouds in which UV photons can be produced by penetration of CRs. However, these mechanisms are still difficult to quantify for astrophysical conditions, and their resulting products may be clusters rather than individual molecules.

Previous research on thermal and non-thermal desorption of molecules from the grain mantle assumed that grains are at rest, which is contrary to the fact that grains are rapidly rotating due to collisions with gas atoms and interstellar photons (\citealt{1998ApJ...508..157D}; \citealt{Hoang:2010jy}). {To accurately understand how molecules are released into the gas}, the effect of grain suprathermal rotation on gas-grain chemistry must be quantified. The goal of this paper is to quantify the effect of grain rotation on the desorption of molecules from icy grain mantles.

Interstellar dust grains are known to be rotating suprathermally, as required to reproduce starlight polarization and far-IR/submm polarized dust emission (see \citealt{Andersson:2015bq} and \citealt{LAH15} for reviews). \cite{1979ApJ...231..404P} first suggested that dust grains can be spun-up to suprathermal rotation (with velocities larger than grain thermal velocity) by various mechanisms, including the formation of hydrogen molecules on the grain surface. Modern astrophysics establishes that dust grains of irregular shapes can rotate suprathermally due to radiative torques arising from their interaction with an anisotropic radiation field (\citealt{1996ApJ...470..551D}; \citealt{2007MNRAS.378..910L}; \citealt{Hoang:2008gb}; \citealt{2009ApJ...695.1457H}; \citealt{Herranen:2019kj}) or mechanical torques induced by an anisotropic gas flow (\citealt{2007ApJ...669L..77L}; \citealt{2018ApJ...852..129H}). As a result, in star-forming regions and photodissociation regions (PDRs), strong radiation can both heat dust grains to high temperatures and spin them up to extremely fast rotation, such that resulting centrifugal force would have an important effect on molecule desorption. 

\cite{Hoang:2019td} first studied the effect of suprathermal rotation induced by radiative torques on the desorption of molecules from the icy grain mantle. For a grain model made of a silicate core covered with a thick ice mantle which is expected in very dense clouds (\citealt{2010ApJ...716..825O}), they discovered that the resulting centrifugal force is sufficient to disrupt the entire ice mantle into small fragments. Subsequently, molecules can evaporate from these fragments due to transient heating by UV photons or enhanced thermal sublimation.\footnote{In this paper, sublimation and desorption is interchangeably used to imply the desorption of molecules from the grain surface.} This process that can desorb the entire ice mantle is then referred to as {\it rotational desorption}. The rotational desorption mechanism is found to be efficient in an extended region beyond hot cores/corinos surrounding young stellar objects (YSOs). Later on, \cite{Le:2019wo} found that molecules can be directly ejected from ice mantles of suprathermally rotating nanoparticles in CJ-shocks.

Another popular grain model consists of a silicate core, an organic refractory layer and outer ice layer (\citealt{1996A&A...309..258G}; \citealt{1997AdSpR..19..981G}; \citealt{2013A&A...558A..62J}). For this model, the ice mantle is presumably thin, of tens of monolayers of ice water, such that it is hard to disrupt the entire ice mantle because the resulting tensile stress is insufficient to separate the binding energy between the mantle and the grain core surface as we will show in Section \ref{sec:theory}. In this case, the joint action of centrifugal force applied to molecules and thermal fluctuations would enhance the rate of thermal sublimation of molecules from the ice mantle, triggering desorption at temperatures below the thermal sublimation threshold. The goal of this paper is to formulate a model of thermal desorption for suprathermally rotating grains and explore its implications for astrochemistry.


The structure of our paper is as follows. In Section \ref{sec:theory}, we first describe the theory of thermal desorption in the presence of grain rotation, which is termed rotational-thermal or {\it ro-thermal desorption}. In Section \ref{sec:result} we calculate the rate of thermal and ro-thermal desorption for grains spun-up by radiation torques from a strong radiation field. Section \ref{sec:discuss} discusses the implications of ro-thermal desorption of molecules and polycyclic aromatic hydrocarbons (PAHs) for different astrophysical environments. A summary of our main findings is presented in Section \ref{sec:sum}. 

\section{Rotational-Thermal Desorption of Molecules from Icy Grain Mantle}\label{sec:theory}
Here we describe our theory for rotational-thermal desorption from rotating grains of angular velocity $\omega$.
\subsection{Grain model: Ice mantles on grain surface}
Ice mantles are formed on the grain surface due to accretion of gas molecules in cold and dense regions of hydrogen density $n_{\H}=n(\H)+2n(\H_{2}) \sim 10^{3}-10^{5}\cm^{-3}$ or the visual extinction $A_{V}> 3$ (\citealt{1983Natur.303..218W}). Subsequently, more complex molecules, including organic molecules, are thought to form in the ice mantle when grains are being warmed up by intense radiation of protostars (see, e.g., \citealt{Herbst:2009go}).

Spectral absorption features of H$_{2}$O and CO ice are highly polarized (\citealt{1996ApJ...465L..61C}; \citealt{2008ApJ...674..304W}) revealing that icy grain mantles have non-spherical shape and are aligned with magnetic fields (see \citealt{LAH15} for a review). {Nevertheless, we assume that the grain shape can be described by an equivalent sphere of the same volume with effective radius $a$.}

\begin{figure}
\centering
\includegraphics[width=0.5\textwidth]{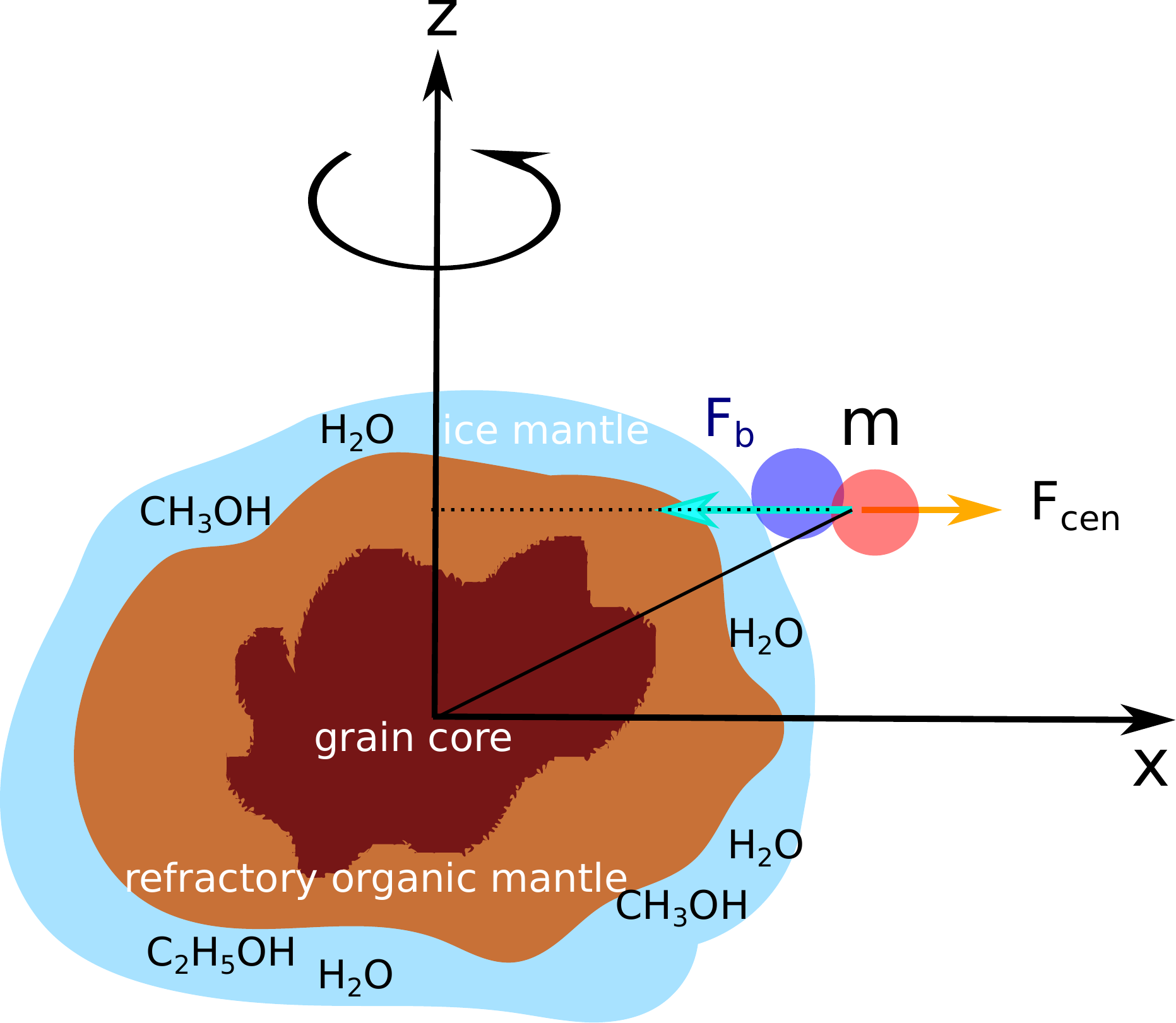}
\caption{A schematic illustration of a rapidly spinning core-mantle grain of irregular shape. The silicate core is assumed to be compact, which is covered with a refractory organic mantle and the outer, icy water-rich mantle. A molecule of mass $m$ on the ice surface experiences the binding force $F_{b}$ and centrifugal force $F_{\rm cen}$ which are in opposite directions.}
\label{fig:grain_mod}
\end{figure}

Figure \ref{fig:grain_mod} illustrates a grain model consisting of a silicate core, followed by a refractory carbonaceous mantle, and an outer thin ice mantle. Molecules bind to the ice mantle via binding force ($F_{b}$) arising from dipole-dipole interaction (van der Waals force) or chemical force. The grain spinning with angular velocity $\omega$ induces a centrifugal force ($F_{cen}$) on the molecule of mass $m$.

\subsection{Thermal desorption rate from non-rotating grains}

The problem of thermal desorption from a non-rotating grain is well studied in the literature (\citealt{1972ApJ...174..321W}; \citealt{1985A&A...144..147L}). The underlying physics is that when the grain is heated to high temperatures, molecules on the grain surface acquire kinetic energy from thermal fluctuations within the grain lattice and can escape from the surface.

Let $\tau_{\rm des,0}$ be the desorption rate of molecules with binding energy $E_{b}$ from a grain at rest ($\omega=0$) which is heated to temperatures $T_{d}$. Following \cite{1972ApJ...174..321W}, one has
\bea
\tau_{\rm sub,0}^{-1}=\nu\exp\left(-\frac{E_{b}}{kT_{d}}\right),
\ena
where $\nu$ is the characteristic frequency given by
\bea
\nu =\left(\frac{2N_{s}E_{b}}{\pi^{2}m}\right)^{1/2}
\ena
with $N_{s}$ being the surface density of binding sites \citep{1987ppic.proc..333T}. Typically, $N_{s}\sim 2\times 10^{15}$ site$\cm^{-2}$.

\begin{figure}
\centering
\includegraphics[width=0.5\textwidth]{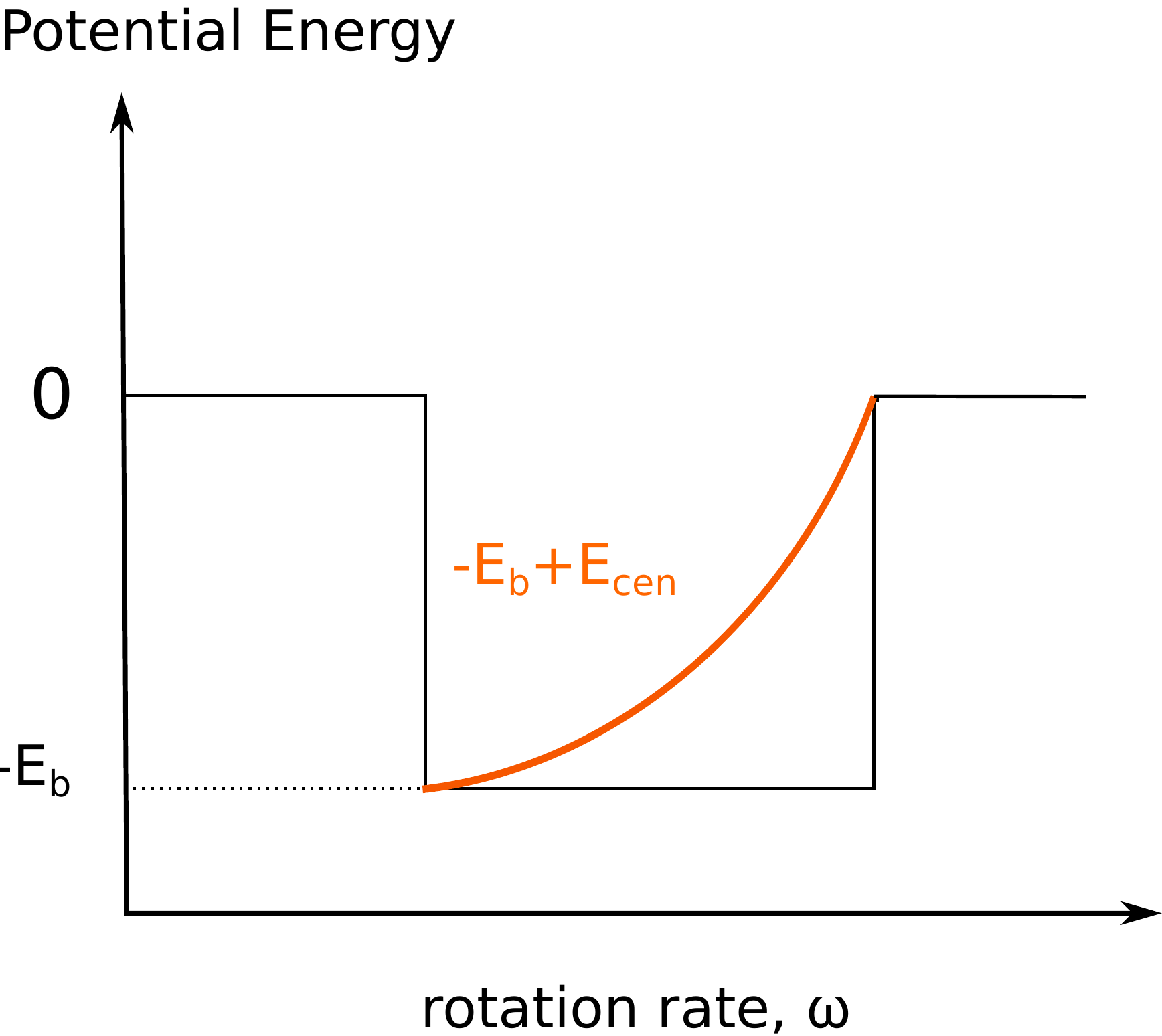}
\caption{Illustration of the potential energy of a molecule on the rotating grain. The potential barrier is reduced significantly as the angular velocity $\omega$ increases as a result of centrifugal potential ($E_{\rm cen}$).}
\label{fig:ETD_mod}
\end{figure}

Table \ref{tab:Ebind} lists the binding energy and sublimation temperature measured from experiments for popular molecules.

\begin{table}
\begin{center}
\caption{Binding energies and sublimation temperatures for selected molecules on an ice surface}\label{tab:Ebind}
\begin{tabular}{l l l} \hline\hline
{Molecules} & {$E_{b}/k$ (K)$^a$} & {$T_{\rm sub}$ (K)}\cr
\hline\\

$\rm H_{2}O$ & 5700 & 152$^b$ \cr
$\rm CH_{3}OH$ & 5530 & 99$^b$ \cr
$\rm HCOOH$ & 5570 & 155$^c$ \cr
$\rm CH_{3}CHO$ & 2775 & 30$^c$ \cr
$\rm C_{2}H_{5}OH$ & 6260 & 250$^c$ \cr
$\rm (CH_{2}OH)_2$ & 10200 & 350$^c$ \cr
$\rm NH_{3}$ & 5530 & 78$^b$\cr
$\rm CO_{2}$ & 2575 & 72$^b$ \cr
$\rm H_{2}CO$ & 2050 & 64$^b$ \cr
$\rm CH_{4}$ & 1300 & 31$^b$ \cr
$\rm CO$ & 1150 & 25$^b$ \cr
$\rm N_{2}$ & 1140 & 22$^b$  \cr

\cr
\hline
\multicolumn{3}{l}{$^a$~See Table 4 in \cite{2013ApJ...765...60G}}\cr
\multicolumn{3}{l}{$^b$~See Table 1 from \cite{1993prpl.conf.1177M}}\cr
\multicolumn{3}{l}{$^c$~See \cite{2004MNRAS.354.1133C}}\cr

\cr
\end{tabular}
\end{center}
\end{table}

\subsection{Ro-thermal desorption rate from rotating grains}

In the presence of grain rotation, the centrifugal force acting on a molecule of mass $m$ at distance $r\sin\theta$ from the spinning axis is
\bea
{\bf F}_{\rm cen}=m{\bf a}_{\rm cen}=m\omega^{2}{\bf r}\sin\theta=m\omega^{2}\sin\theta(x\xhat+y\yhat),\label{eq:Fcen}
\ena
where $\ba_{\rm cen}$ is the centrifugal acceleration.

We can define centrifugal potential $\phi_{\rm cen}$ such as ${\bf a}_{\rm cen}=-\nabla\phi_{\rm cen}$. Then, the corresponding potential is 
\bea
\phi_{\rm cen}=\omega^{2}\sin^{2}\theta\left(\frac{x^{2}+y^{2}}{2}\right)=\frac{1}{2}\omega^{2}\sin^{2}\theta r^{2}.
\ena
{Assuming that molecules are uniformly distributed over the grain surface, then, one can obtain the average centrifugal potential as follows:}
\bea
\langle \phi_{\rm cen}\rangle=\frac{\omega^{2}a^{2}\langle \sin^{2}\theta\rangle}{2}=\frac{\omega^{2}a^{2}}{3},\label{eq:phi_cen}
\ena
{where $\langle \sin^{2}\theta\rangle=2\int_{0}^{\pi/2} \sin^{2}\theta \sin\theta d\theta=2/3$.}

As a result, the {\it effective} binding energy of the molecule becomes
\bea
E_{b,rot}=E_{b}-m\langle \phi_{\rm cen}\rangle,\label{eq:Ebind_rot}
\ena
which means that molecules only need to overcome the reduced potential barrier of $E_{b}-E_{\rm cen}$ where $E_{\rm cen}=m\langle\phi_{\rm cen}\rangle$ to be ejected from the grain surface. The rotation effect is more important for molecules with higher mass and low binding energy.

Figure \ref{fig:ETD_mod} illustrates the potential barrier of molecules on the surface of a rotating grain as a function of $\omega$. For slow rotation, the potential barrier is determined by binding force. As $\omega$ increases, the potential barrier is decreased due to the contribution of centrifugal potential.

{The molecule is instantaneously ejected from the surface if the grain is spinning sufficiently fast such that $E_{b,rot}=0$. From Equation (\ref{eq:Ebind_rot}), one can obtain the critical angular velocity for the direct ejection as follows}:
\bea
\omega_{\rm ej}=\left(\frac{3E_{b}}{ma^{2}}\right)^{1/2}\simeq \frac{10^{10}}{a_{-5}}\left(\frac{(E_{b}/k)}{1300\K}\frac{m_{\rm CO}}{m}\right)^{1/2}\rm rad\s^{-1},~~~\label{eq:omega_ej}
\ena
where $a_{-5}=a/(10^{-5}\cm)$.

The ejection angular velocity decreases with increasing grain size and molecule mass $m$, but it increases with the binding energy $E_{b}$. 

The rate of ro-thermal desorption (sublimation) rate is given by
\bea
\tau_{\rm sub,rot}^{-1}=\nu\exp\left(-\frac{E_{b}-m\langle \phi_{\rm cen}\rangle}{kT_{d}}\right),\label{eq:tsub_rot}
\ena
where {the subscript $\rm sub$ stands for sublimation}, and the second exponential term describes the probability of desorption induced by centrifugal potential.

Equation (\ref{eq:tsub_rot}) can be written as
\bea
\tau_{\rm sub,rot}^{-1}=\tau_{\rm sub,0}^{-1}RD(\omega),
\ena
where the function $RD(\omega)$ describes the effect of grain rotation on the thermal desorption as given by
\bea
RD(\omega)&=&\exp\left(\frac{m\langle \phi_{\rm cen}\rangle}{kT_{d}} \right)=\exp\left(\frac{m\omega^{2}a^{2}}{3kT_{d}} \right)\\
&&\simeq1.7\exp\left[a_{-5}^{2}\left(\frac{m}{m_{\rm CO}}\right)\left(\frac{\omega}{10^{9}\s^{-1}}\right)^{2}\left(\frac{20\K}{T_{d}}\right)\right]\nonumber
\ena
which indicates the rapid increase of ro-thermal desorption rate with the grain size $a$ and angular velocity $\omega$.

Let $\tilde{\omega}=\omega/\omega_{T}$ be the suprathermal rotation parameter where $\omega_{T}=(2kT/I)^{1/2}\simeq 2\times 10^{5}a_{-5}^{-5/2}T_{2}^{1/2}\rad\s^{-1}$ with $T$ gas temperature and $T_{2}=T/100\K$, and $I=8\pi \rho a^{5}/15$ inertia moment of grains with mass volume density $\rho$. Then, one obtains
\bea
RD(\omega)=\exp\left(\frac{2m a^{2}}{3I}\frac{T}{T_{d}} \tilde{\omega}^{2}\right)=\exp\left(\frac{5m}{3M}\frac{T}{T_{d}} \tilde{\omega}^{2}\right)
\ena
where $M=4\pi \rho a^{3}/3$ is the grain mass.

\begin{figure*}
\includegraphics[width=0.5\textwidth]{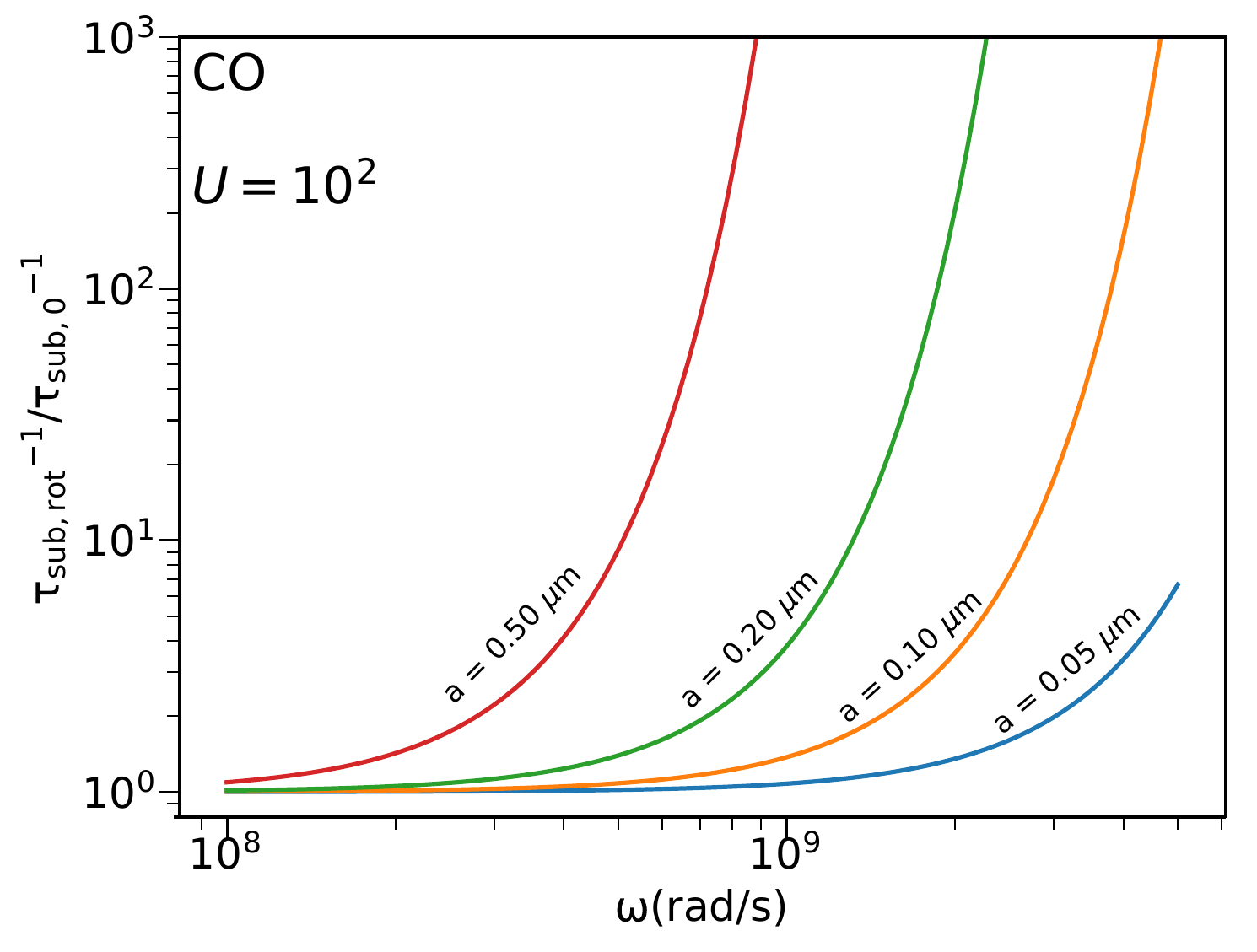}
\includegraphics[width=0.5\textwidth]{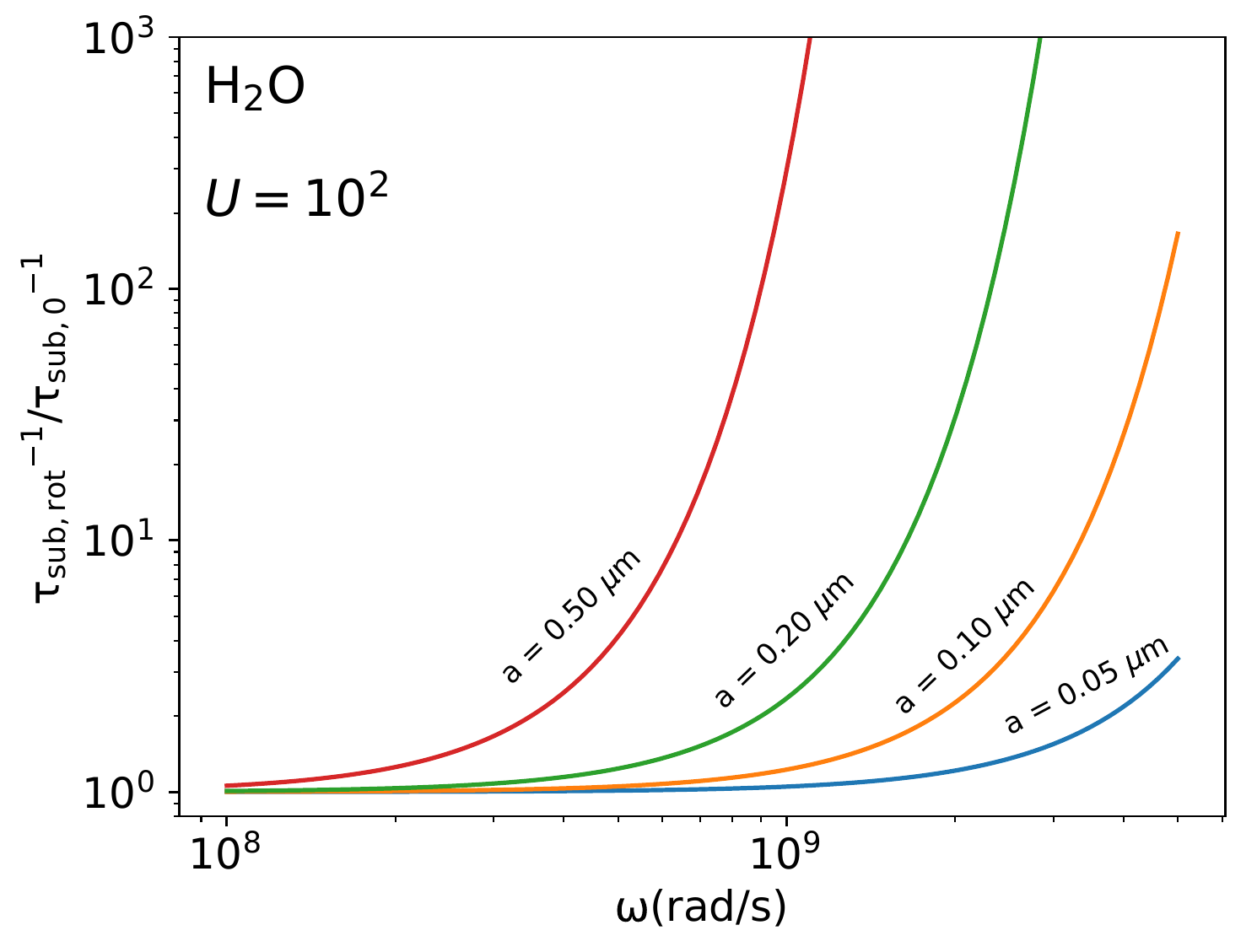}
\includegraphics[width=0.5\textwidth]{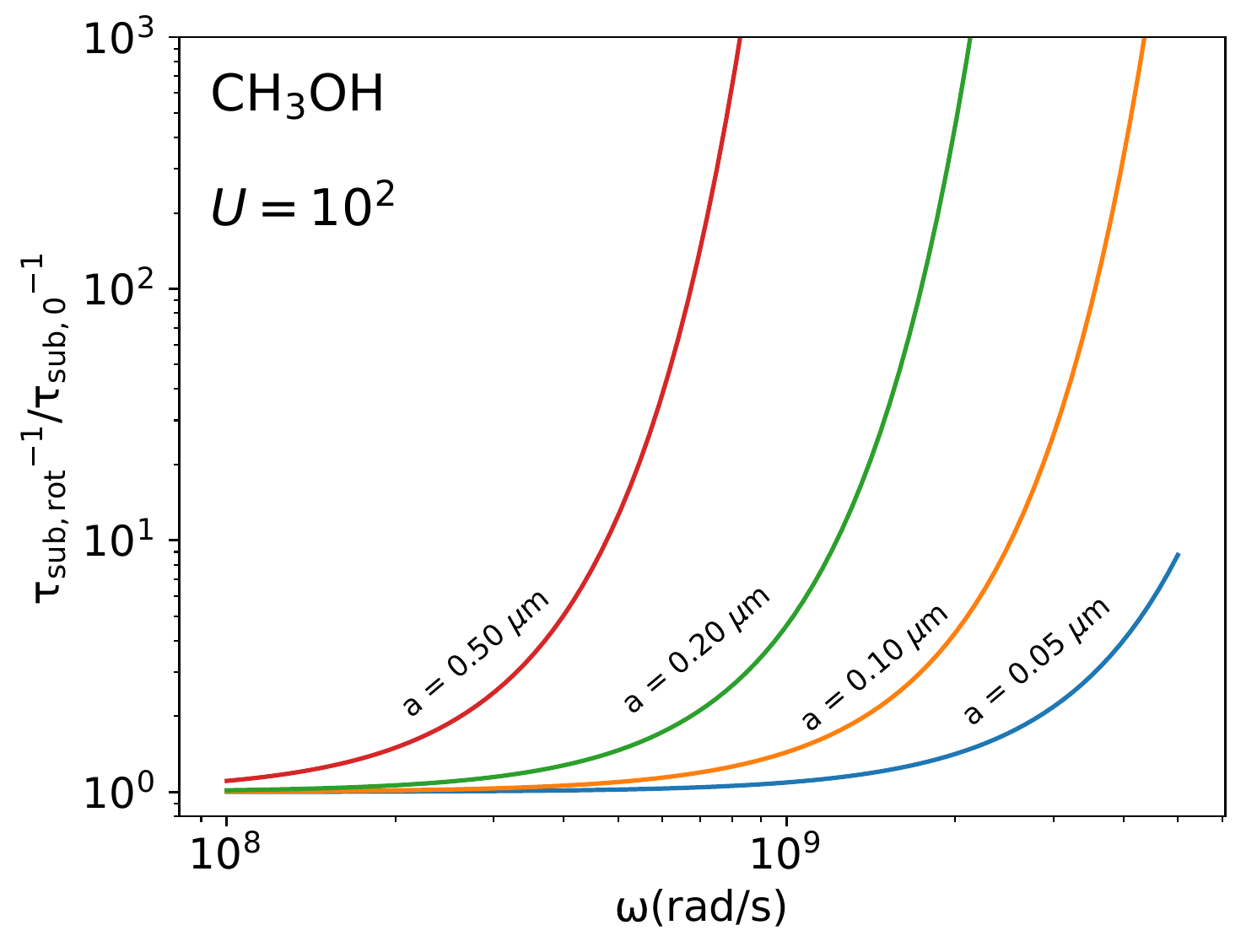}
\includegraphics[width=0.5\textwidth]{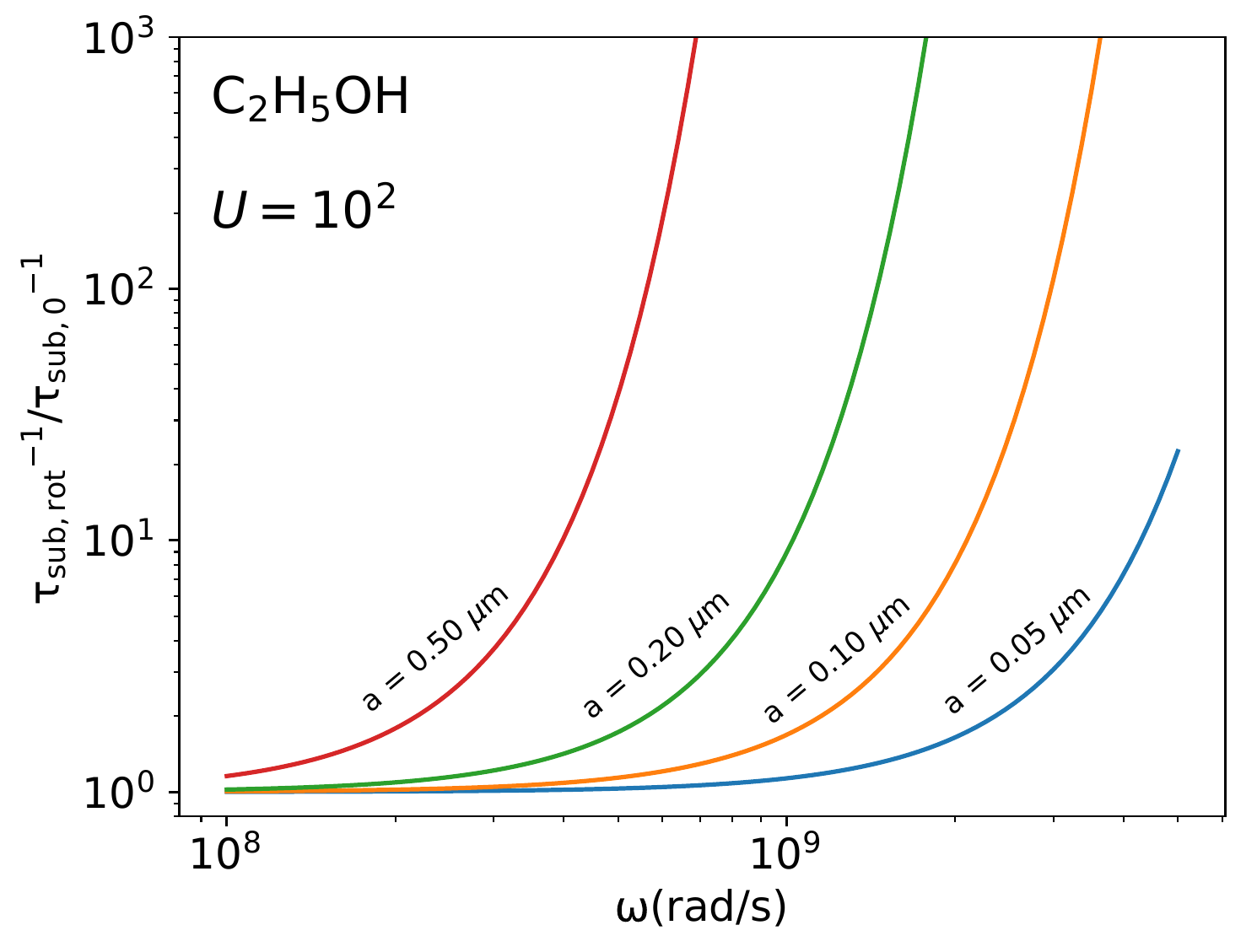}
\caption{The ratio of ro-thermal desorption rate to thermal desorption rate as a function of grain angular velocity $\omega$ computed for the different molecules and different grain sizes, assuming the grain temperature of $T_{d}\approx 35.5\K$ ($U=100$). The ro-thermal desorption efficiency increases with increasing $\omega$ and the grain size $a$.}
\label{fig:frot_omega}
\end{figure*}

Figure \ref{fig:frot_omega} shows the ratio of ro-thermal to thermal sublimation rate, $RD(\omega)$, as a function of the grain angular velocity. For a given grain temperature, the rate of ro-thermal desorption increases exponentially with the angular velocity $\omega$ when $\omega$ is approaching $\omega_{\rm ej}$ (Eq. \ref{eq:omega_ej}).

\subsection{Sublimation temperatures from rotating grains}
Let $T_{\rm sub,0}$ be the sublimation temperature of grains at rest, i.e., $\omega=0$. The sublimation temperature of rotating grains is denoted by $T_{\rm sub,rot}$.
To quantify the effect of grain rotation on thermal desorption, we compare the grain temperature that is required to produce the same sublimation rate from a non-rotating grain which corresponds to $\tau_{\rm sub,0}(T_{\rm sub,0})=\tau_{\rm sub,rot}(T_{\rm sub,rot})$. Thus, one obtains
\bea
T_{\rm sub,rot}=\left(1-\frac{m\langle \phi_{\rm cen}\rangle}{E_{b}} \right)T_{\rm sub,0}.\label{eq:Tsub_rot}
\ena

The effect of grain rotation reduces the sublimation temperature as given by
\bea
\frac{T_{\rm sub,0}-T_{\rm sub,rot}}{T_{\rm sub,0}}&=&\left(\frac{m\langle \phi_{\rm cen}\rangle}{E_{b}}\right)=\left(\frac{ma^{2}\omega^{2}}{3E_{b}}\right)\label{eq:dTsub}\\
&\simeq& 0.14a_{-5}^{2}\left(\frac{\omega}{5\times 10^{9}\s^{-1}}\right)^{2}\left(\frac{m}{m_{\rm CO}}\right)\left(\frac{2000\K}{(E_{b}/k)}\right).\nonumber
\ena

One can see that the the ro-thermal sublimation temperature can be decreased by $50\%$ for grains of $a=0.2\mum$ rotating at $\omega=5\times 10^{9}\rad\s^{-1}$.

\section{Ro-thermal desorption from grains spun-up by radiative torques}\label{sec:result}
{In this section, we will apply the theory formulated in the preceding section for the situation in which grains are rotating suprathermally as a result of the spin-up by Radiative Torques (RATs, e.g., \citealt{1996ApJ...470..551D}).}
\subsection{Centrifugal potential due to radiative torques}
Following \cite{Hoang:2019bi}, subject to a radiation field of anisotropy degree $\gamma$, mean wavelength $\bar{\lambda}$, and radiation strength $U$, dust grains of size $a$ can be spun-up to a maximum rotation rate by (RATs):
\bea
\omega_{\rm RAT}&\simeq &9.6\times 10^{8}\gamma a_{-5}^{0.7}\bar{\lambda}_{0.5}^{-1.7}\nonumber\\
&\times&\left(\frac{U}{n_{1}T_{2}^{1/2}}\right)\left(\frac{1}{1+F_{\rm IR}}\right)\rad\s^{-1},~~~\label{eq:omega_RAT1}
\ena
for grains with $a\lesssim \bar{\lambda}/1.8$, and
\bea
\omega_{\rm RAT}&\simeq &1.78\times 10^{10}\gamma a_{-5}^{-2}\bar{\lambda}_{0.5}\nonumber\\
&&\times\left(\frac{U}{n_{1}T_{2}^{1/2}}\right)\left(\frac{1}{1+F_{\rm IR}}\right)\rad\s^{-1},~~~\label{eq:omegaRAT2}
\ena
for grains with $a> \overline{\lambda}/1.8$. Here, $n_{1}=n_{\H}/(10\cm^{-3}), \bar{\lambda}_{0.5}=\bar{\lambda}/(0.5\mum)$, $F_{\rm IR}$ is the dimensionless parameter describing the grain rotational damping by infrared emission that depends on $(n_{\H}, T_{\rm gas}, U)$ (\citealt{1998ApJ...508..157D}; \citealt{Hoang:2010jy}), and $U=u_{\rad}/u_{\rm ISRF}$ with $u_{\rad}$ the total radiation energy density and $u_{\rm ISRF}$ the energy density of the standard interstellar radiation field (ISRF) in the solar neighborhood (\citealt{1983A&A...128..212M}; \citealt{Hoang:2019da}). The rotation rate depends on the parameter $U/n_{\H}T_{\gas}^{1/2}$ and the damping by far-infrared emission $F_{\rm IR}$. 

For convenience, let $a_{\rm trans}=\bar{\lambda}/1.8$ which denotes the grain size at which the RAT efficiency changes between the power law and flat stages (see e.g., \citealt{2007MNRAS.378..910L}; \citealt{Hoang:2019da}), and $\omega_{\rm RAT}$ changes from Equation (\ref{eq:omega_RAT1}) to (\ref{eq:omegaRAT2}). 

Plugging $\omega_{\rm RAT}$ into Equation (\ref{eq:phi_cen}), one obtains the centrifugal potential due to grain rotation as follows:
\bea
m\langle \phi_{\rm cen}\rangle&=&1.8\times 10^{-3}\gamma^{2} a_{-5}^{3.4}\bar{\lambda}_{0.5}^{-3.4}\left(\frac{m}{m_{\rm CO}}\right)\nonumber\\
&\times&\left(\frac{U}{n_{1}T_{2}^{1/2}}\right)^{2}\left(\frac{1}{1+F_{\rm IR}}\right)^{2}~\rm eV\label{eq:phi_small}
\ena
for $a\lesssim a_{\rm trans}$ and
\bea
m\langle \phi_{\rm cen}\rangle&=&0.6\gamma^{2} a_{-5}^{-2}\bar{\lambda}_{0.5}^{2}\left(\frac{m}{m_{\rm CO}}\right)\nonumber\\
&\times&\left(\frac{U}{n_{1}T_{2}^{1/2}}\right)^{2}\left(\frac{1}{1+F_{\rm IR}}\right)^{2} \rm eV\label{eq:phi_big}
\ena
for $a>a_{\rm trans}$.

The centrifugal potential increases rapidly with the grain size as $a^{3.4}$ until $a=a_{\rm trans}$ (Eq.\ref{eq:phi_small}), and it increases with the radiation strength as $U^{2}$. Thus, this potential is important for strong radiation fields. 

Using the centrifugal potentials (Eqs. \ref{eq:phi_small} and \ref{eq:phi_big}) one can calculate the rate of ro-thermal desorption (Eq. \ref{eq:tsub_rot}) and the temperature threshold for ro-thermal desorption (Eq. \ref{eq:Tsub_rot}).

\subsection{Radiation strength required for rotational desorption of molecules}
In addition to the ro-thermal desorption, individual molecules can be directly ejected by centrifugal forces when the rotational rate is sufficiently high. This process is termed {\it rotational desorption} in \cite{Le:2019wo}. Comparing $\omega_{\rm RAT}$ with $\omega_{\rm ej}$ (Eq. \ref{eq:omega_ej}), one can then derive the critical radiation strength at which the molecule is immediately ejected
\bea
U_{\rm ej}\simeq 8n_{1}T_{2}^{1/2}(1+F_{\rm IR})\frac{\lambda_{0.5}^{1.7}}{\gamma a_{-5}^{1.7}}\left(\frac{(E_{b}/k)}{1300\K}\frac{m_{\rm CO}}{m}\right)^{1/2}~\label{eq:Uej}
\ena
for $a\lesssim a_{\rm trans}$, and 
\bea
U_{\rm ej}\simeq 0.4n_{1}T_{2}^{1/2}(1+F_{\rm IR})\frac{\lambda_{0.5}^{1.7} a_{-5}}{\gamma}\left(\frac{(E_{b}/k)}{1300\K}\frac{m_{\rm CO}}{m}\right)^{1/2}
\ena
for $a> a_{\rm trans}$.

\subsection{Radiation strength required for rotational desorption of entire ice mantles}
As shown in \cite{Hoang:2019td}, when the rotation rate is sufficiently high such as the tensile stress acting on the interface between a thick ice mantle and the grain core exceeds the maximum limit of the ice mantle, $S_{\rm max}\sim 10^{7}\erg\cm^{-3}$, the ice mantle is disrupted into smaller fragments. In the case of a thin ice mantle, the tensile strength is replaced by the adhesive strength, which depends on the mechanical property of the surface and grain temperature. The adhesive strength is low for clean surface, but it can reach $\sim 10^{9}\erg\cm^{-3}$ for rough surfaces \citep{Work:2018bu}. 

Let $l_{m}$ be the ice mantle thickness and $x_{0}$ be the distance from the core-mantle interface to the spinning axis. From Equation (7) in \cite{Hoang:2019td}, one obtains the tensile stress on the ice mantle:
\bea
S_{x}\simeq 2.5\times 10^{9}\hat{\rho}_{\rm ice}\omega_{10}^{2}a_{-5}^{2} \left[1-\left(\frac{x_{0}}{a}\right)^{2} \right] \erg \cm^{-3},\label{eq:Sx}
\ena
where $\hat{\rho}_{\rm ice}=\rho_{\rm ice}/(1\g\cm^{-3})$ and $\omega_{10}=\omega/(10^{10}\rm rad\s^{-1})$. For a thin mantle layer of $l_{m}=a-x_{0}\ll a$, one has
\bea
S_{x}\simeq 5 \times 10^{8}\hat{\rho}_{\rm ice}\omega_{10}^{2}a_{-5}l_{-6}\erg \cm^{-3},\label{eq:Sx_approx}
\ena
where $x_{0}+a\approx 2a$ is assumed, and $l_{-6}=l_{m}/(10^{-6}\cm)$. The critical rotational velocity is determined by $S_{x}=S_{\rm max}$:
\bea
\omega_{\rm disr}&=&\frac{2}{a(1-x_{0}^{2}/a^{2})^{1/2}}\left(\frac{S_{\max}}{\rho_{\rm ice}} \right)^{1/2}\nonumber\\
&\simeq& \frac{4.5\times 10^{10}}{a_{-5}^{1/2}l_{-6}^{1/2}}\hat{\rho}_{\rm ice}^{-1/2}S_{\max,9}^{1/2}~\rad\s^{-1},\label{eq:omega_disr}
\ena
where $S_{\max,9}=S_{\max}/(10^{9} \erg \cm^{-3})$. 

The critical radiation strength to disrupt the ice mantle is then
\bea
U_{\rm disr}\simeq 32.5 n_{1}T_{2}^{1/2}\frac{(1+F_{\rm IR})}{l_{-6}^{1/2}}\left(\frac{\lambda_{0.5}^{1.7}}{\gamma a_{-5}^{1.2}}\right)S_{\rm max,9}^{1/2}\label{eq:Udisr}
\ena
for $a\lesssim a_{\rm trans}$, and 
\bea
U_{\rm disr}\simeq 1.8n_{1}T_{2}^{1/2}\frac{(1+F_{\rm IR})}{l_{-6}^{1/2}}\left(\frac{\lambda_{0.5}^{1.7}a_{-5}^{1.5}}{\gamma}\right)S_{\rm max,9}^{1/2}
\ena
for $a> a_{\rm trans}$.

For grains with a compact core, the tensile strength $S_{\max}\sim 10^{9}\erg\cm^{-3}$ is expected (see e.g., \citealt{Hoang:2019da}). Comparing $\omega_{\rm ej}$ from Equation (\ref{eq:Uej}) with $\omega_{\rm disr}$ from Equation (\ref{eq:Udisr}), one can see that ro-thermal desorption of individual molecules can occur before the disruption of ice mantle if the ice mantle thickness is below 100 monolayers of water ice (i.e., $l_{m}<200$~\AA).

\subsection{Radiation strength required for thermal desorption}
Under strong radiation fields, icy grains are heated to an equilibrium temperature, which can be approximately given by $T_{d}\simeq 16.4 a_{-5}^{-1/15}U^{1/6}\K$ for silicate-core grains (\citealt{2011piim.book.....D}). One can then derive the radiation strength required for the classical thermal sublimation at $T_{d}=T_{\rm sub,0}$:
\bea
U_{\rm sub}\simeq 6\times 10^{11}a_{-5}^{6/15}\left(\frac{T_{\rm sub,0}}{1500\K}\right)^{6}.\label{eq:Usub}
\ena

Comparing $U_{\rm sub}$ with $U_{\rm ej}, U_{\rm disr}$ one can see that the required radiation strength for thermal desorption is many orders of magnitude higher than ro-thermal desorption as well as direct ejection. One note that, for the same radiation strength $U$, graphite grains can be heated to temperatures higher than silicates by $\sim 30\%$, but their sublimation threshold is more than two times higher than silicates. As a result, the value of $U_{\rm sub}$ for graphite grains is much larger.

\subsection{Numerical results}
\subsubsection{Rates of ro-thermal vs. thermal desorption}
To calculate the rate of ro-thermal desorption, we first compute the rotation rate spun-up by RATs as given by Equations (\ref{eq:omega_RAT1}) and (\ref{eq:omegaRAT2}). Here we adopt $\gamma=0.7$ for the anisotropy degree of the radiation field as in \cite{1996ApJ...470..551D} for molecular clouds (see also \citealt{2007ApJ...663.1055B}).\footnote{The anisotropy degree of the diffuse ISRF is lower of $\gamma=0.1$, and $\gamma=1$ for unidirectional radiation fields from a point source.} We then calculate the ro-thermal desorption rate of different molecules from the surface of spinning dust grains. We consider the different gas density and radiation strengths and assume thermal equilibrium between gas and dust, i.e.,  $T=T_{d}$ which is valid for dense regions around protostars. Our calculations are performed for several popular molecules, including methanol, ethanol, with binding energy listed in Table \ref{tab:Ebind}.

\begin{figure*}
\includegraphics[width=0.5\textwidth]{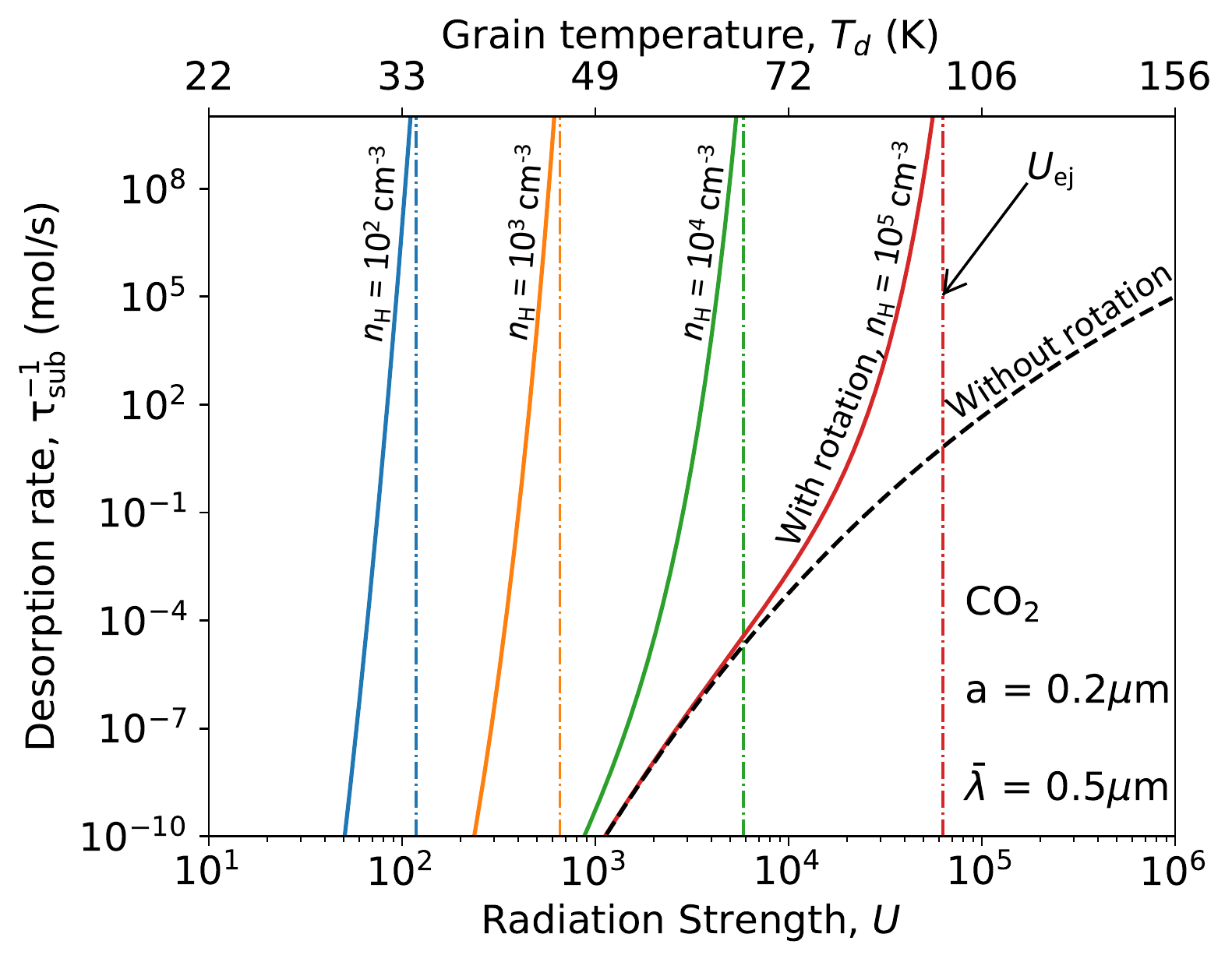}
\includegraphics[width=0.5\textwidth]{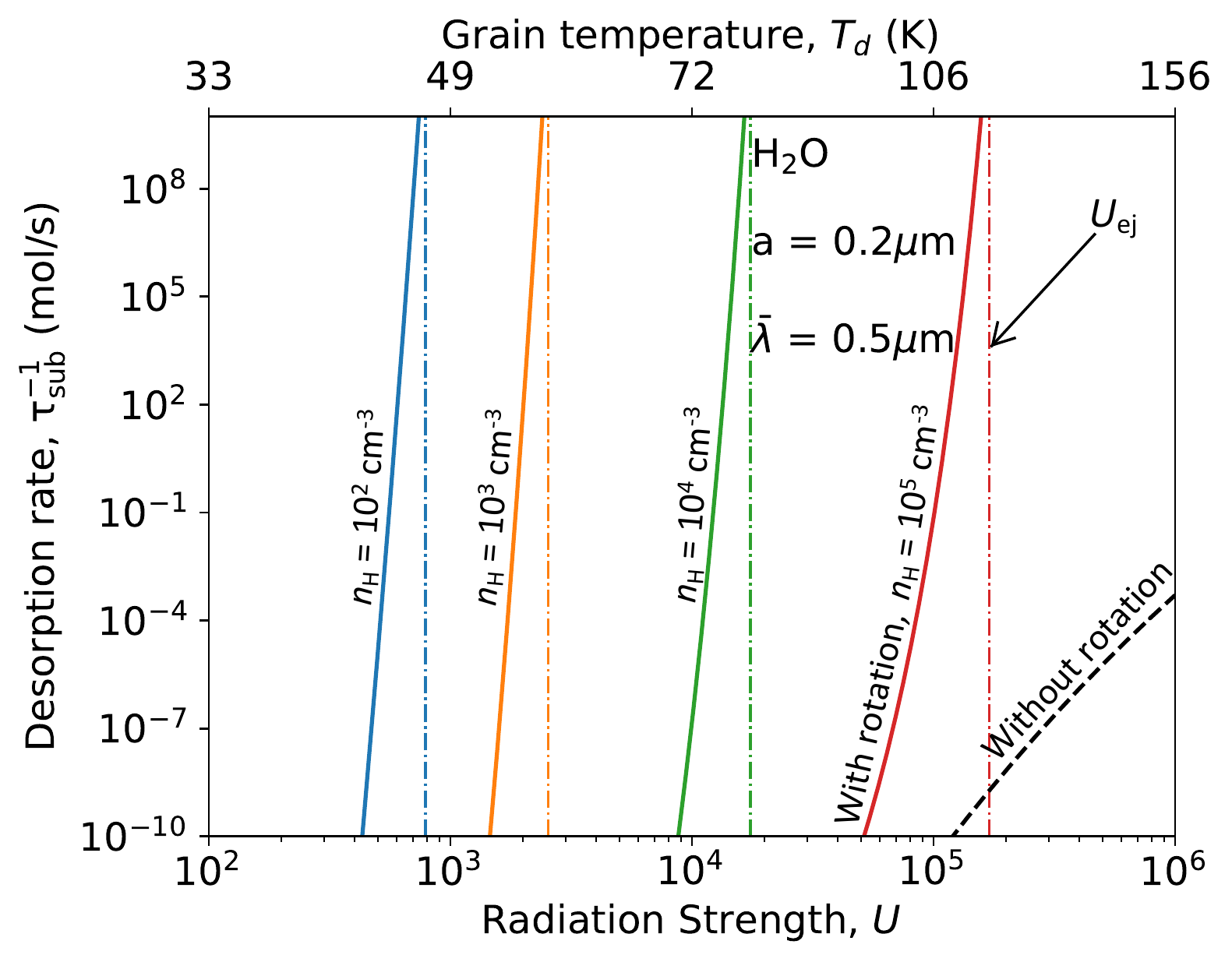}
\includegraphics[width=0.5\textwidth]{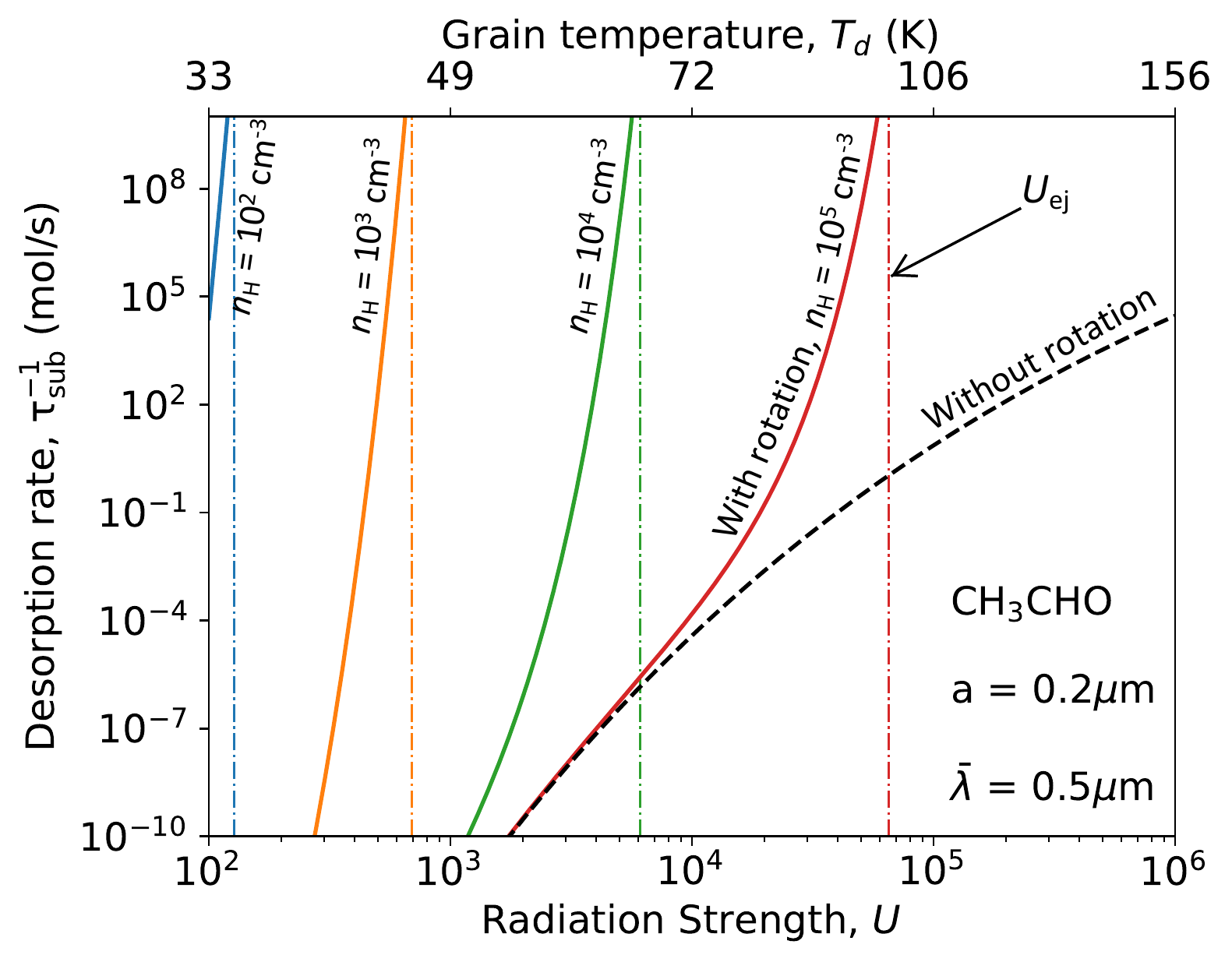}
\includegraphics[width=0.5\textwidth]{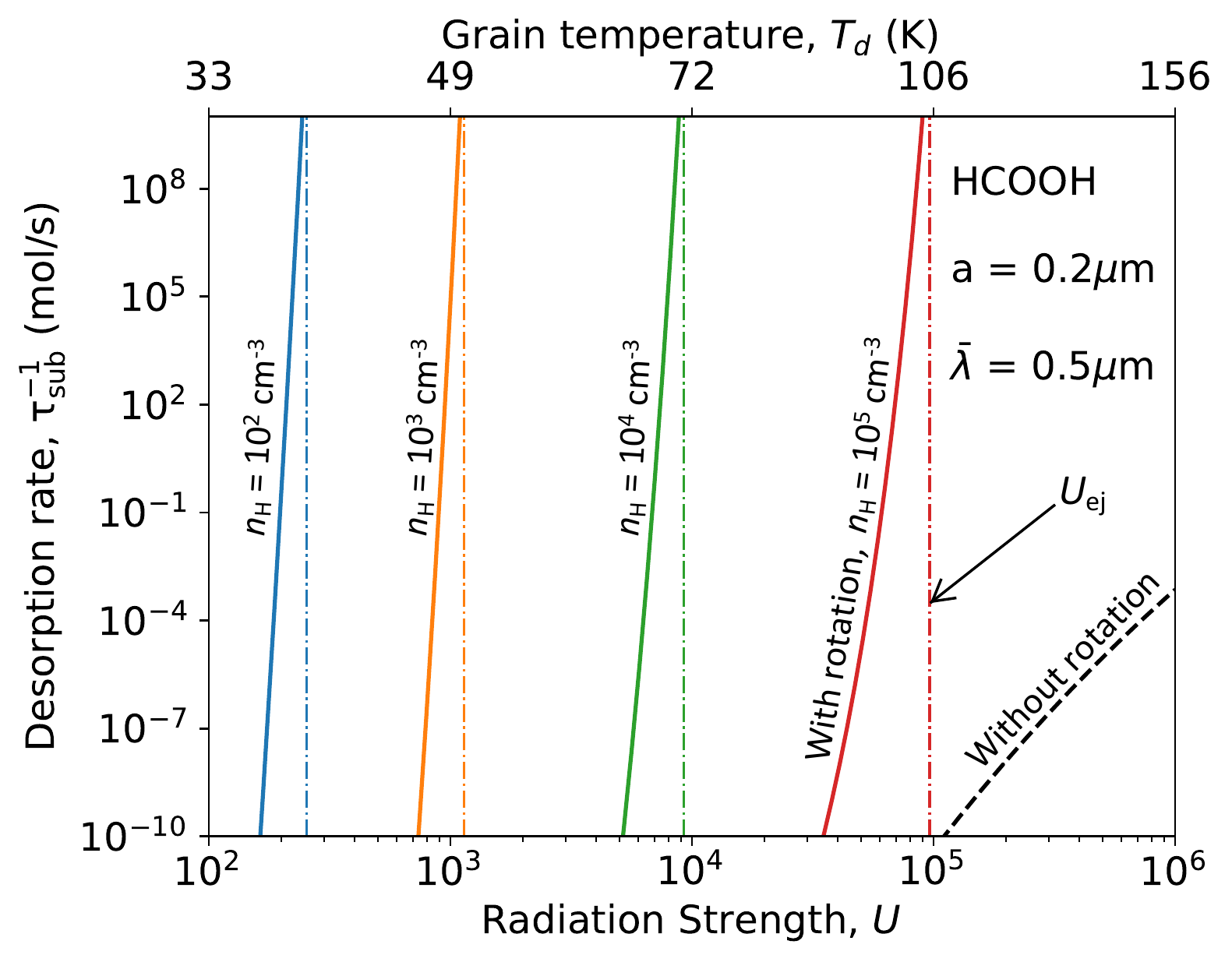}
\includegraphics[width=0.5\textwidth]{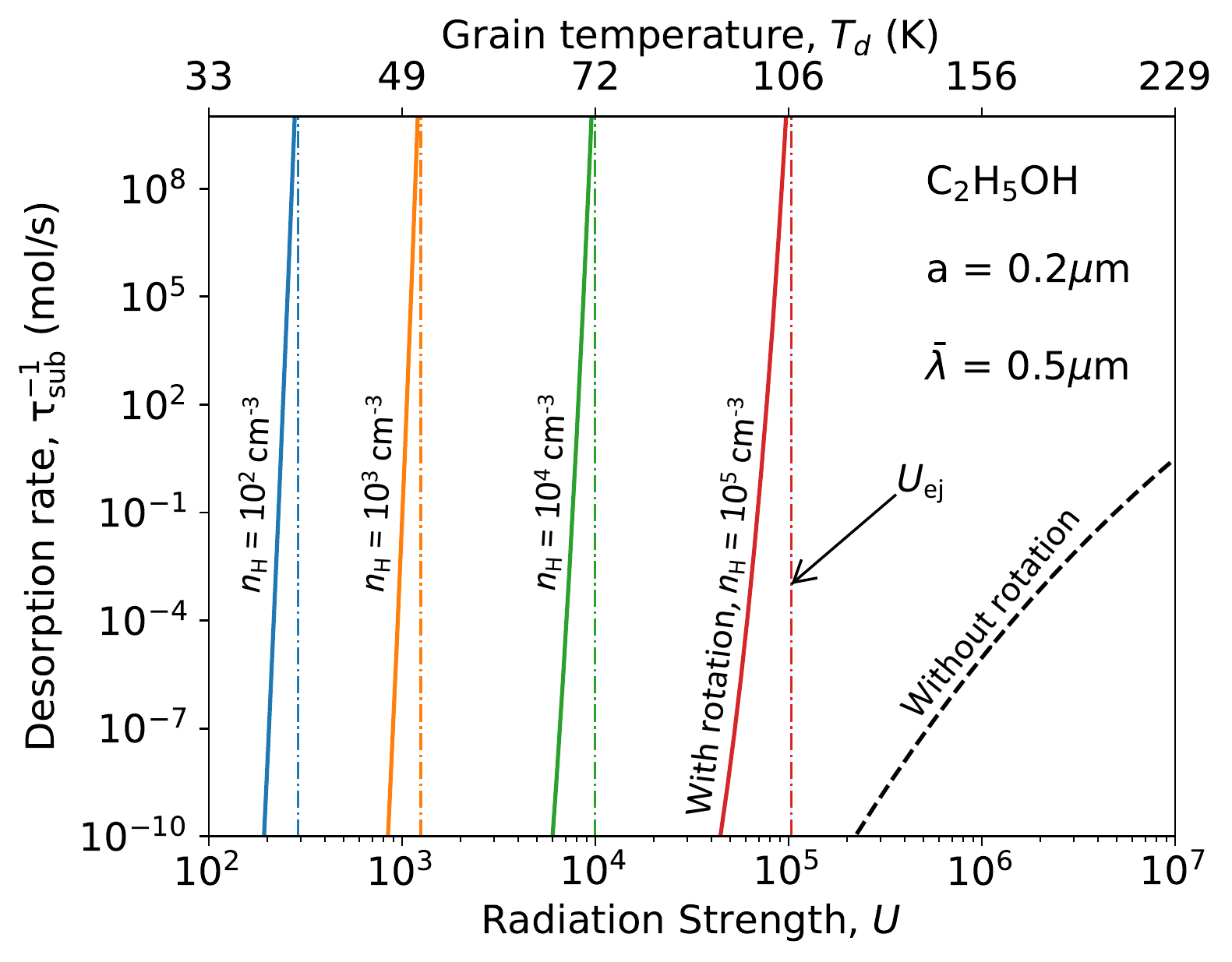}
\includegraphics[width=0.5\textwidth]{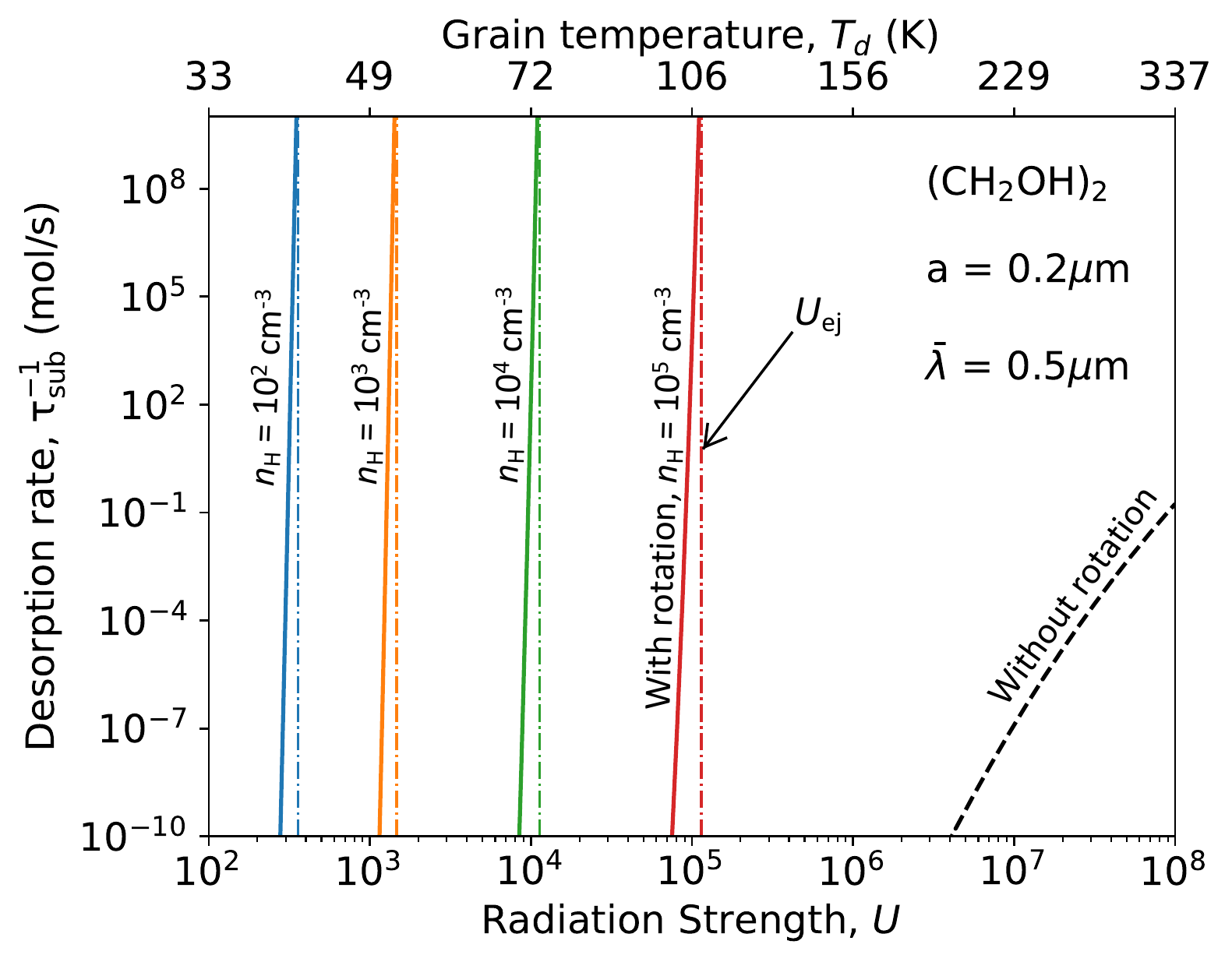}
\caption{Rates of ro-thermal desorption (solid lines, with rotation) and thermal desorption (dashed line, without rotation) for several molecules as function of the radiation strength (U) and grain temperature (top horizontal axis), assuming grains of size $a=0.2\mum$ and the mean wavelength $\bar{\lambda}=0.5\mum$. A range of gas density $n_{\rm H}=10^{2}-10^{5}\cm^{-3}$ is considered. The vertical lines show the direct ejection threshold $U_{\rm ej}$.}
\label{fig:rate_U_wave05_a02}
\end{figure*}

Figure \ref{fig:rate_U_wave05_a02} shows the rate of thermal desorption (without rotation) and ro-thermal desorption (with rotation) as a function of the radiation strength $U$ assuming a typical grain size $a=0.2\mum$ and stellar radiation spectrum with $\bar{\lambda}=0.5\mum$. The corresponding grain temperatures are shown on the top horizontal axis. The ro-thermal desorption rate increases exponentially with the radiation intensity even at temperatures much below the sublimation threshold. In all realizations, the ro-thermal desorption is much faster than thermal desorption except for $CO_{2}$ with high density $n_{\rm H}=10^{5}\cm^{-3}$. The efficiency of ro-thermal desorption is stronger for lower gas density $n_{\rm H}$. This originates from the fact that grains can spin faster due to lower rotational damping by gas collisions. One can also see that for most molecules the ro-thermal desorption occurs well before the immediate ejection threshold marked by $U_{\rm ej}$.

\begin{figure*}
\includegraphics[width=0.5\textwidth]{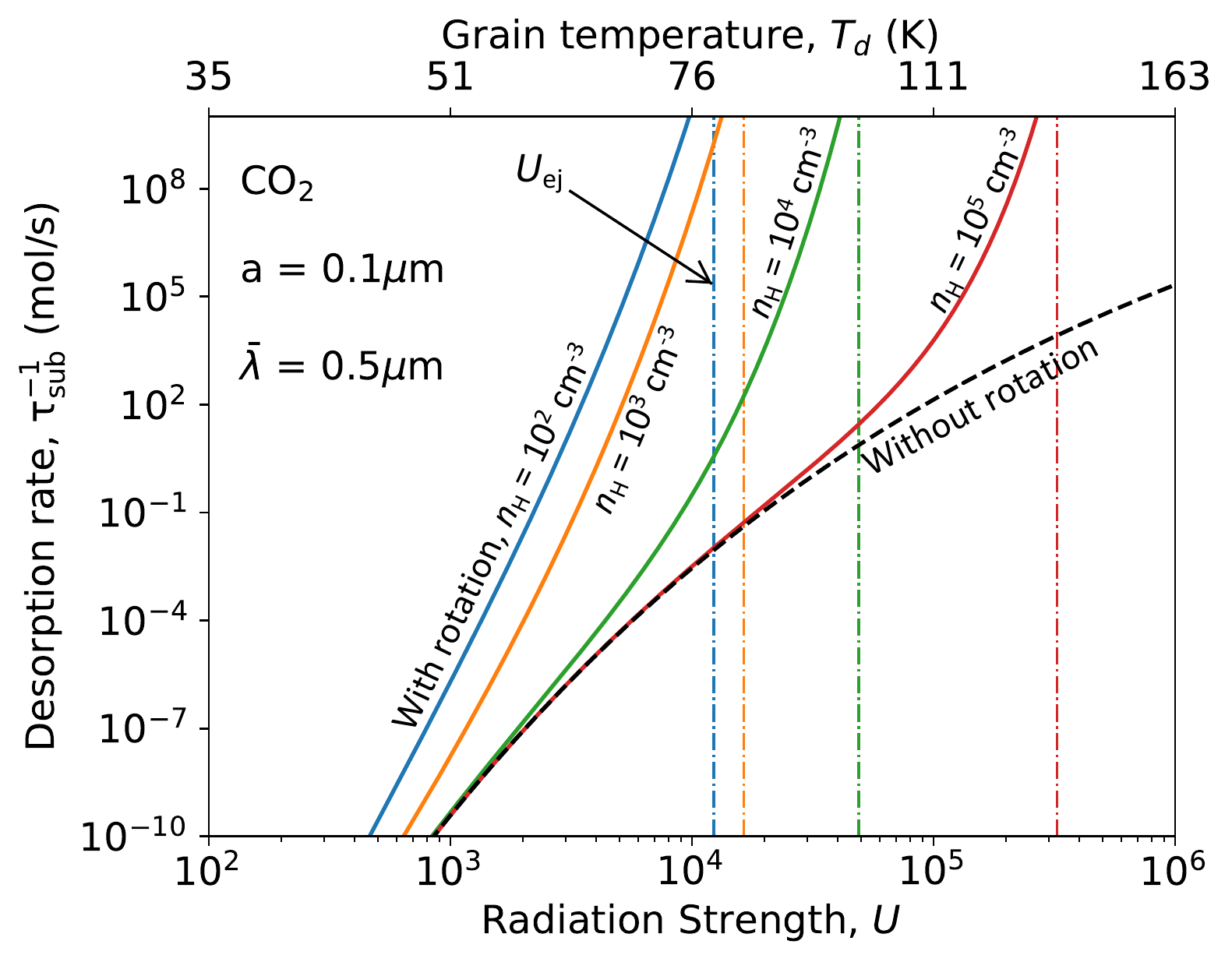}
\includegraphics[width=0.5\textwidth]{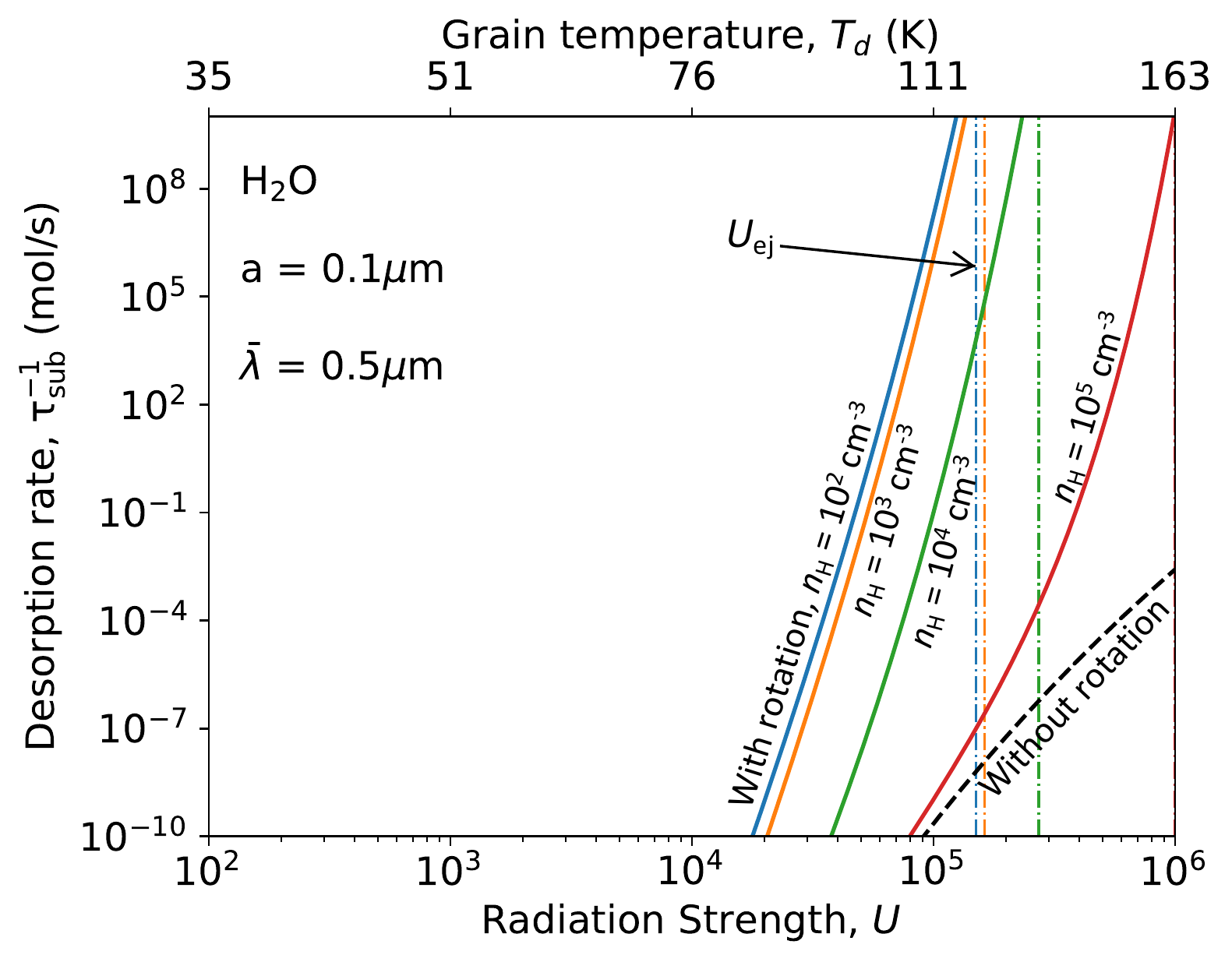}
\includegraphics[width=0.5\textwidth]{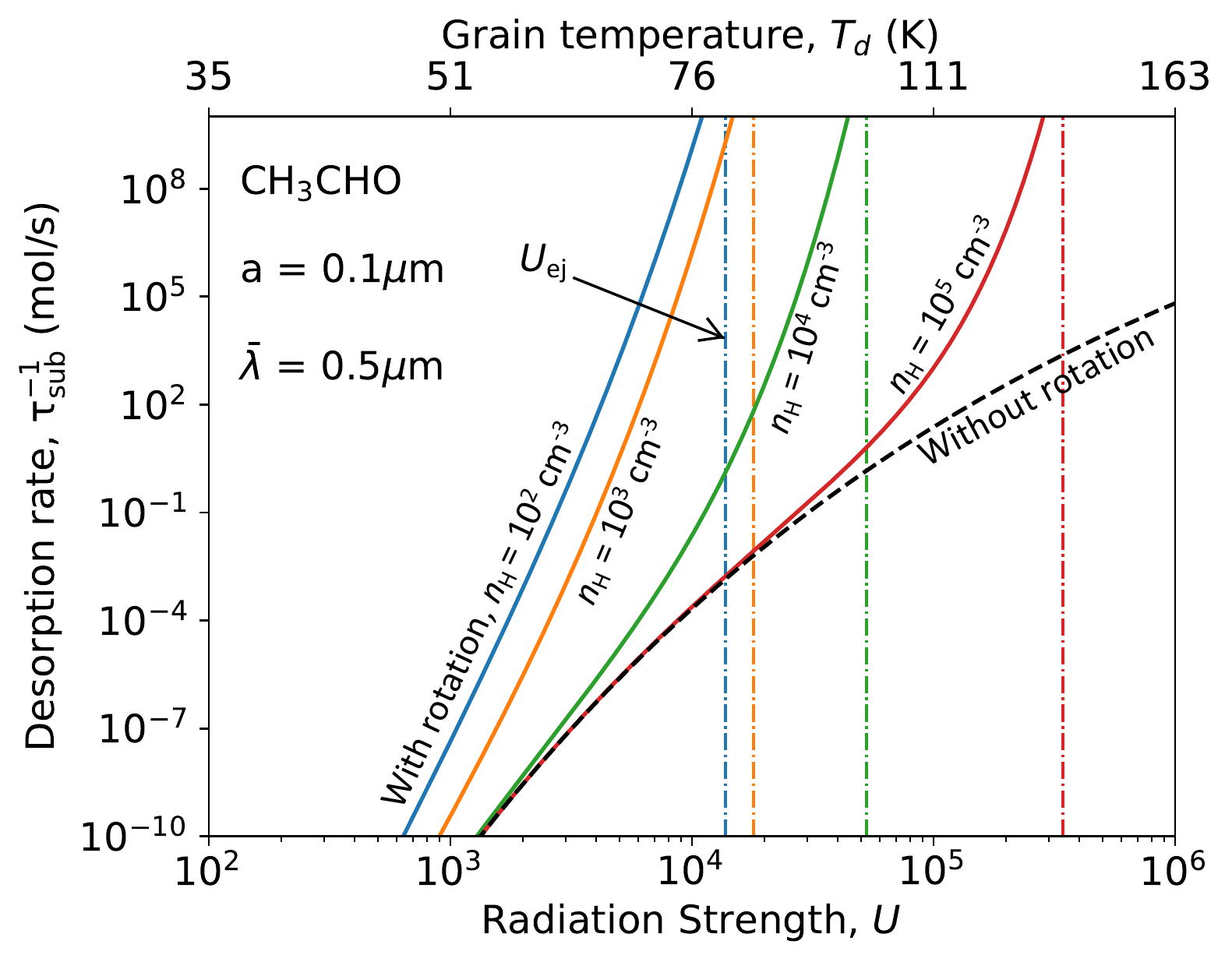}
\includegraphics[width=0.5\textwidth]{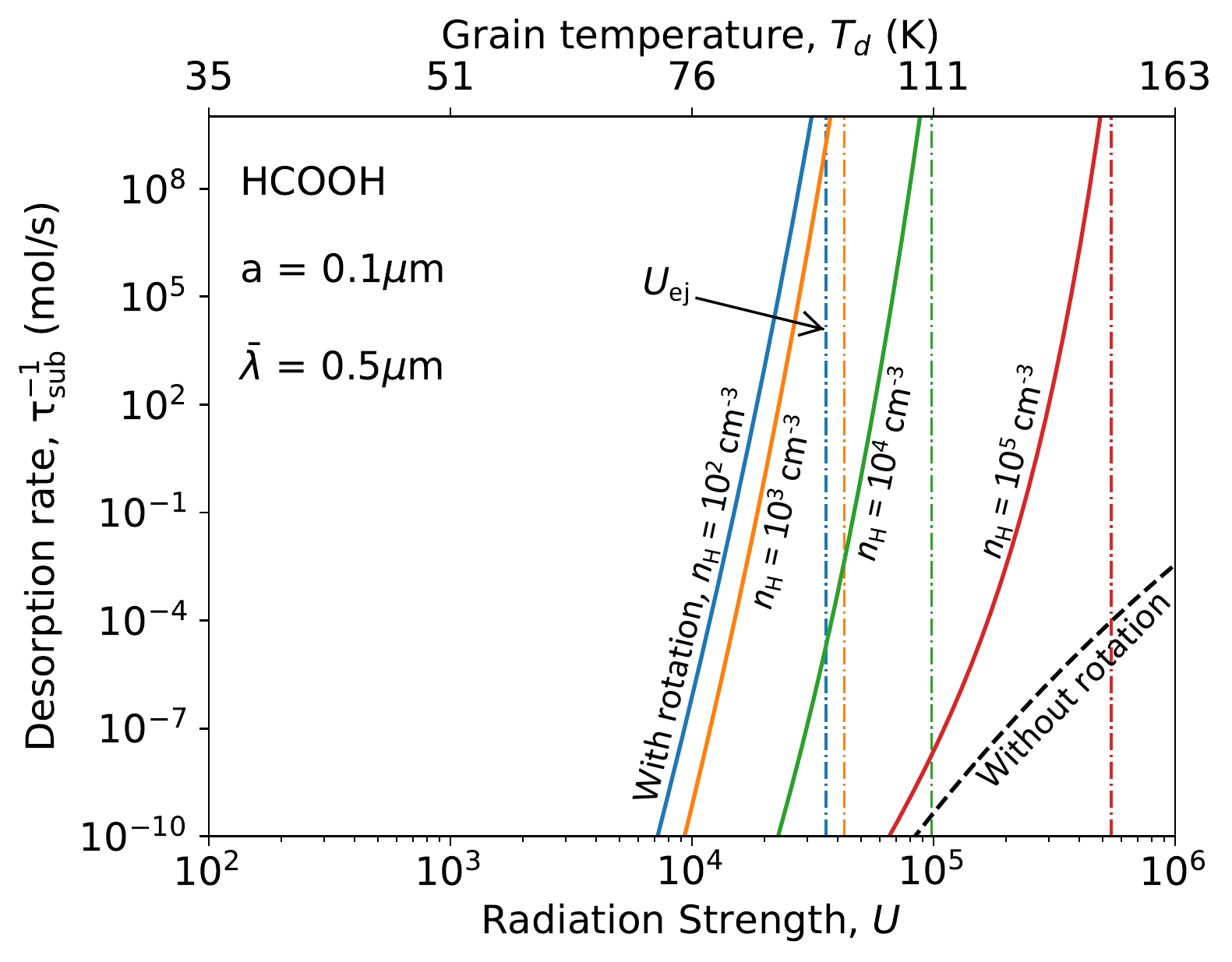}
\includegraphics[width=0.5\textwidth]{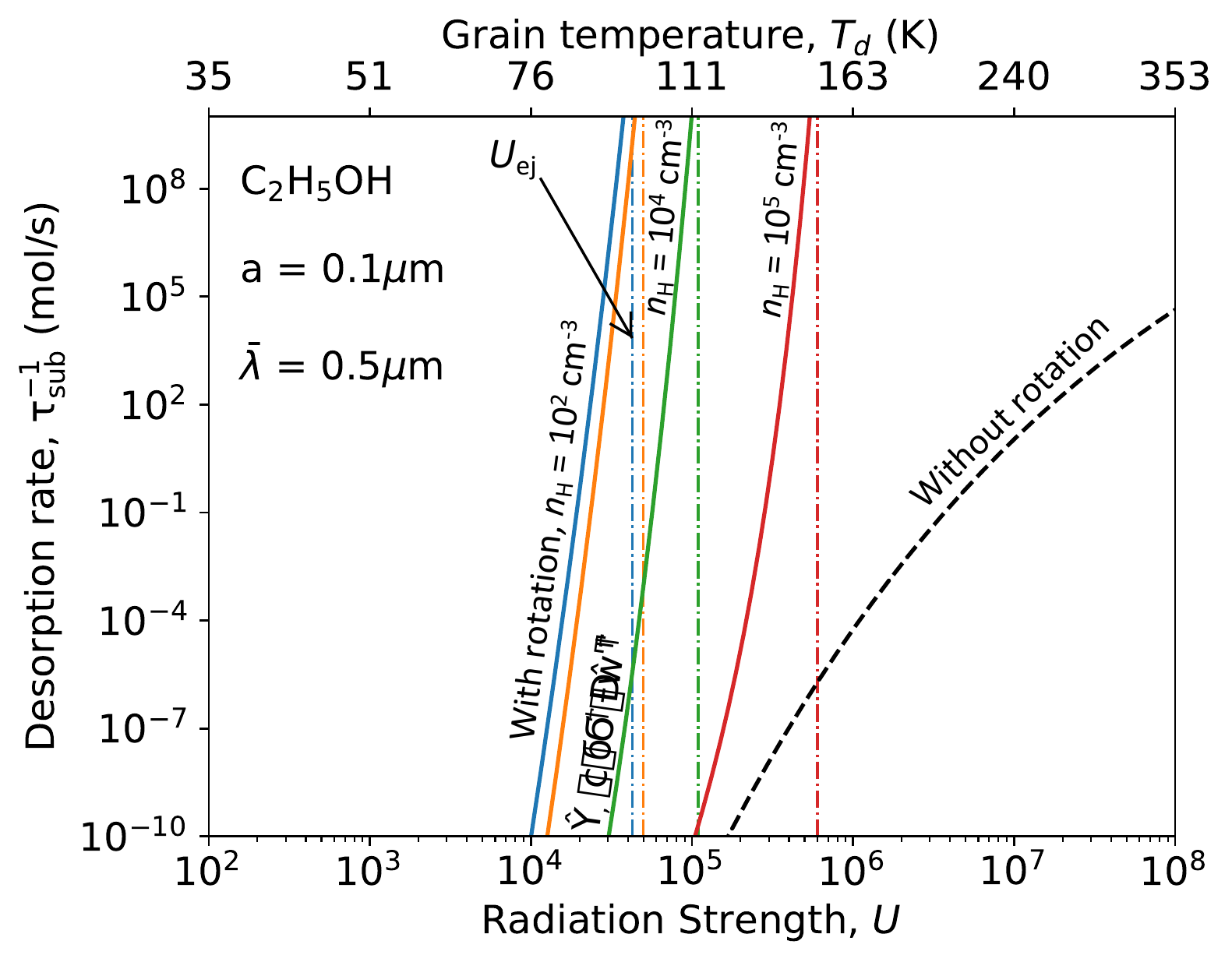}
\includegraphics[width=0.5\textwidth]{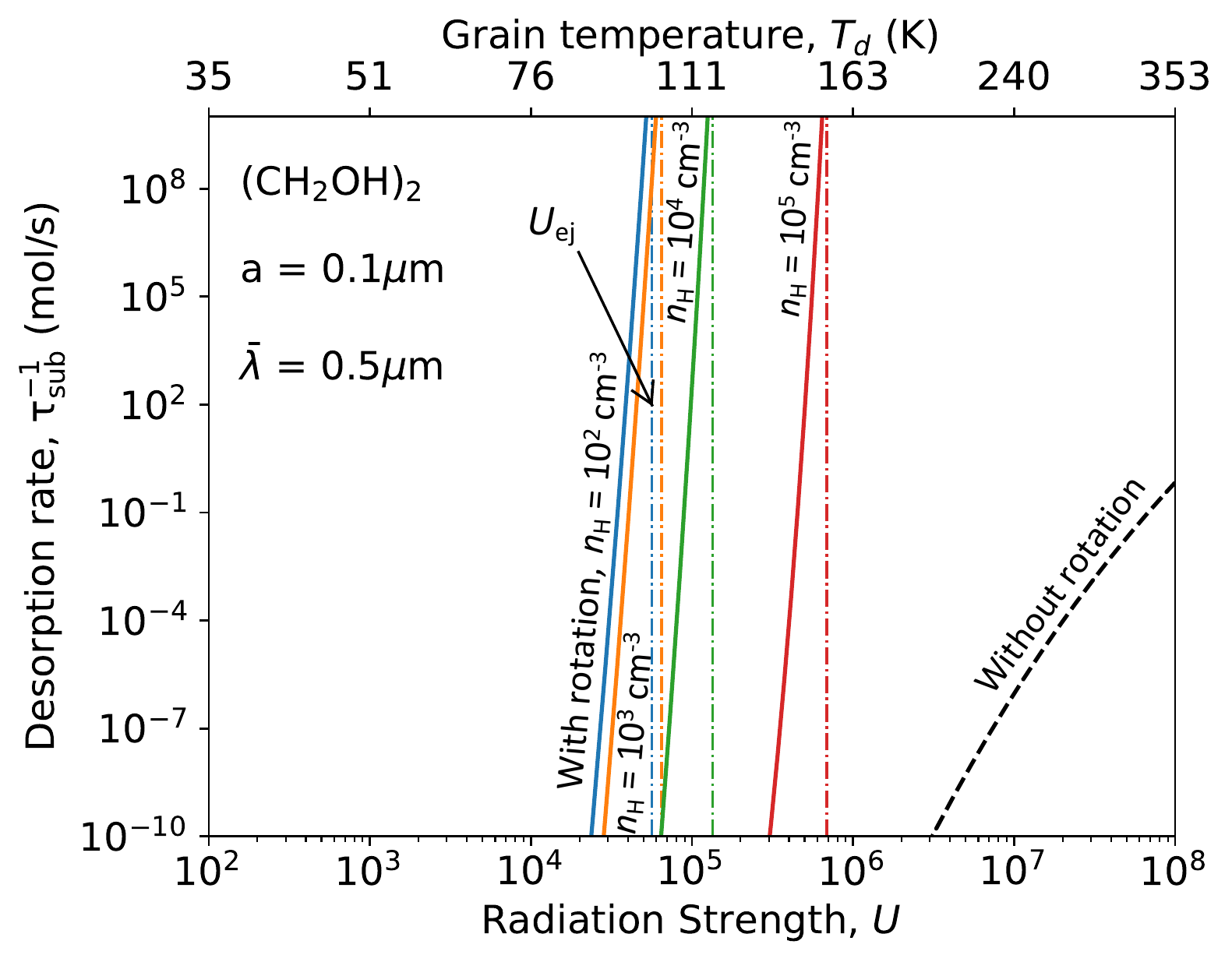}
\caption{Same as Figure \ref{fig:rate_U_wave05_a02} but for grains of size of $a=0.1\mum$. Due to smaller grain size, the efficiency of ro-thermal desorption is decreased but still dominates over thermal desorption (dashed line).}
\label{fig:rate_U_wave05_a01}
\end{figure*}

Figure \ref{fig:rate_U_wave05_a01} shows the rate of ro-thermal vs. thermal desorption for $a=0.1\mum$. Ro-thermal desorption is still faster than thermal desorption, although the efficiency is lower that for $a=0.2\mum$ due to lower rotation rate by RATs $\omega_{\rm RAT}$ (see Eq. \ref{eq:omega_RAT1}).

\begin{figure*}
\includegraphics[width=0.5\textwidth]{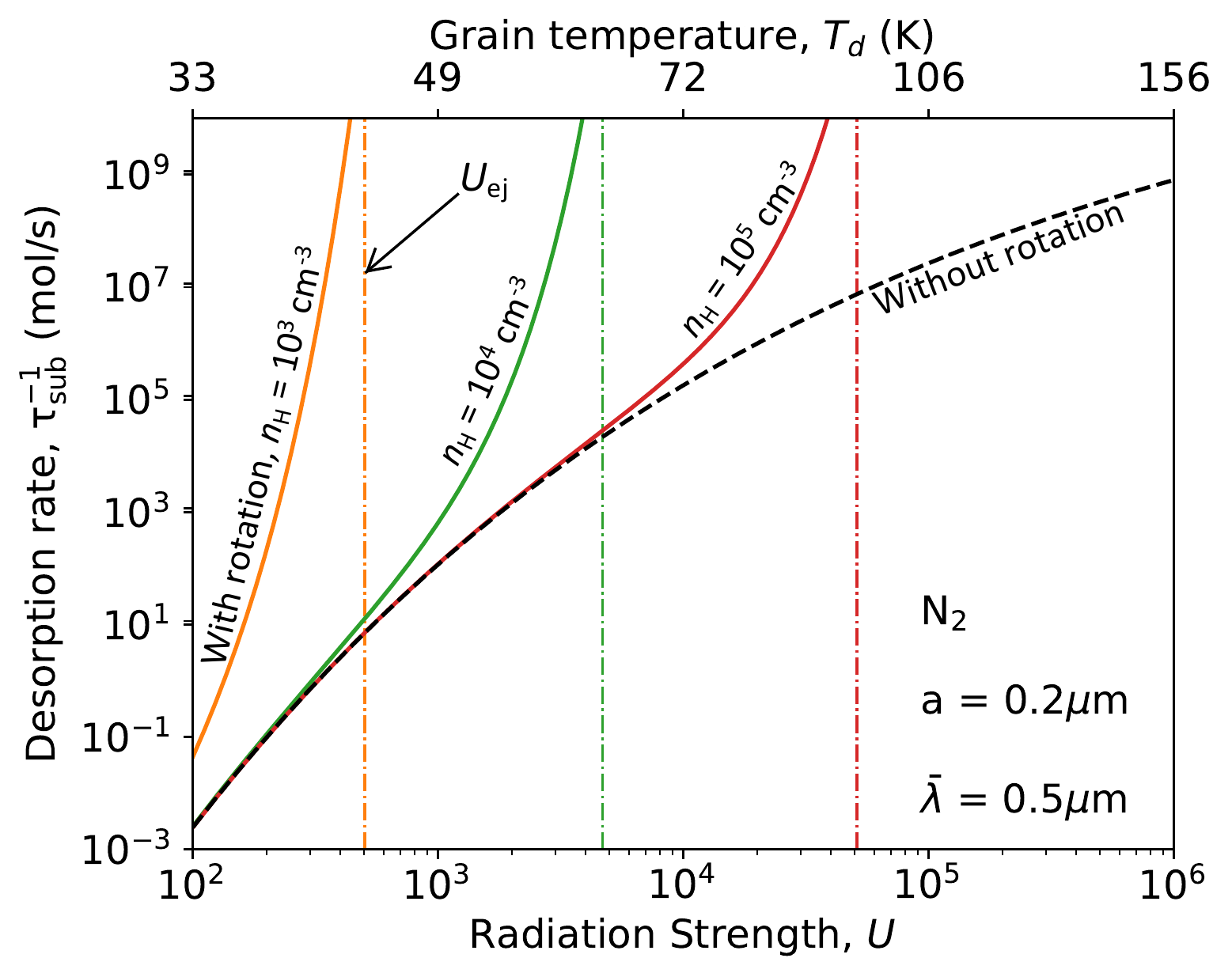}
\includegraphics[width=0.5\textwidth]{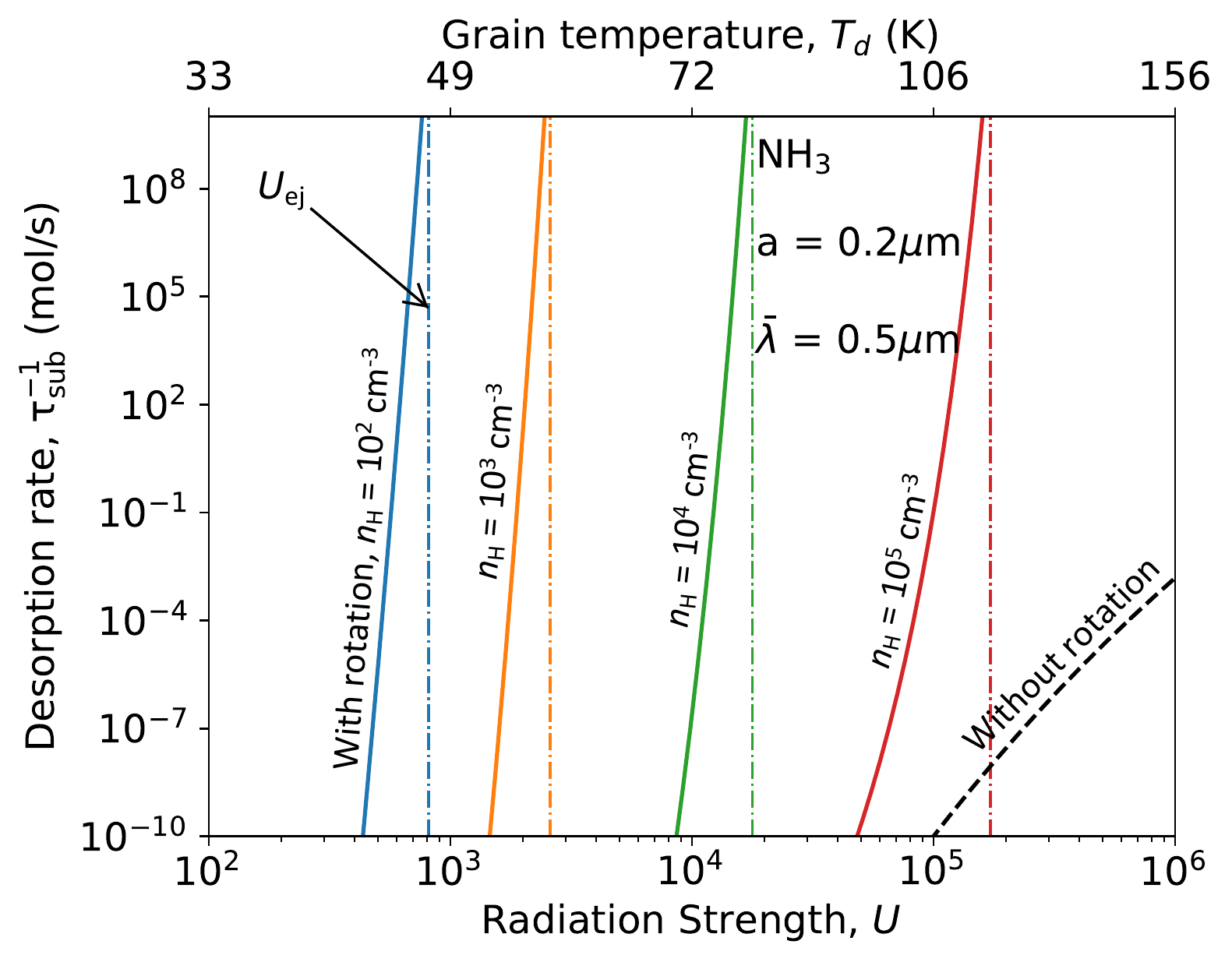}
\includegraphics[width=0.5\textwidth]{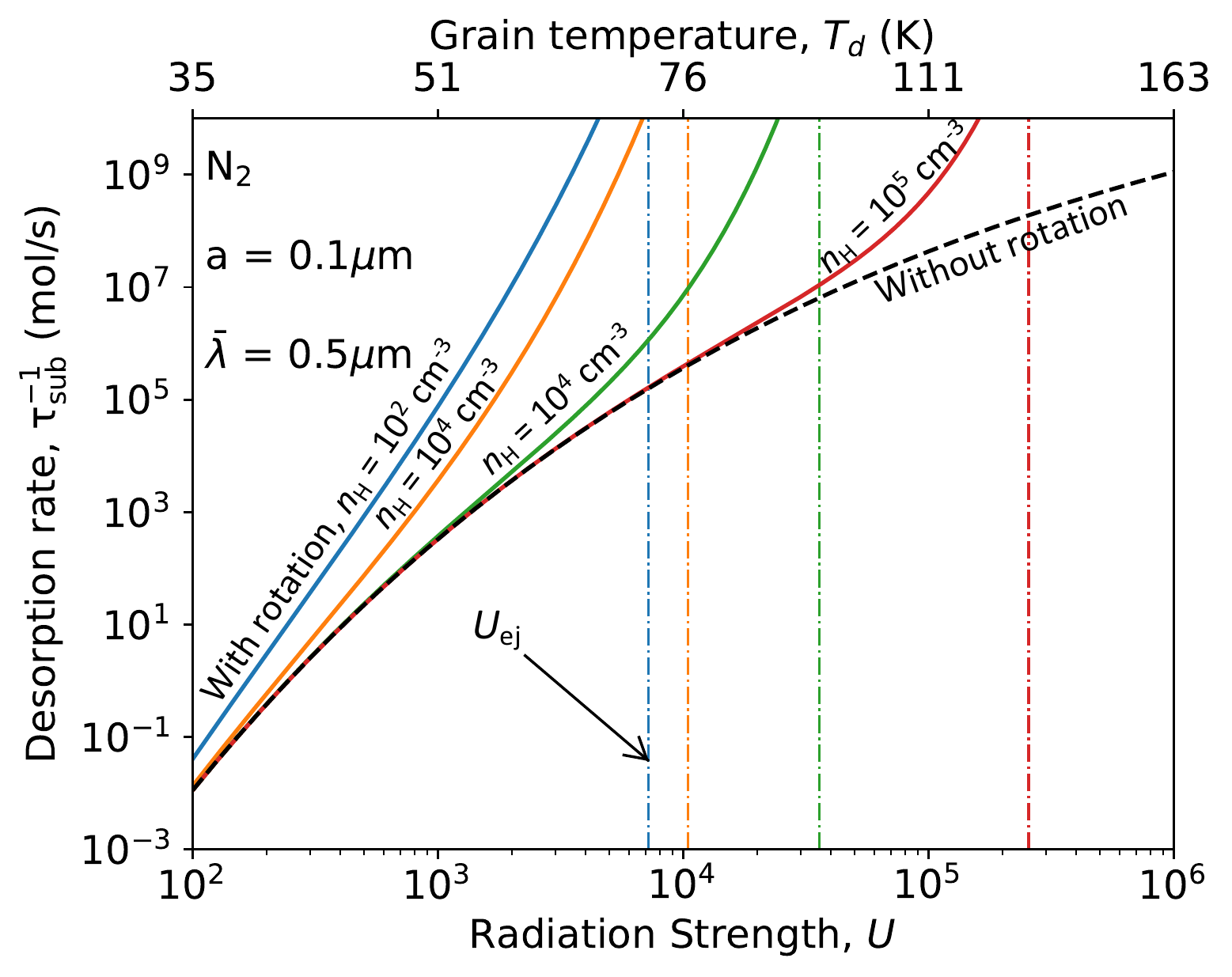}
\includegraphics[width=0.5\textwidth]{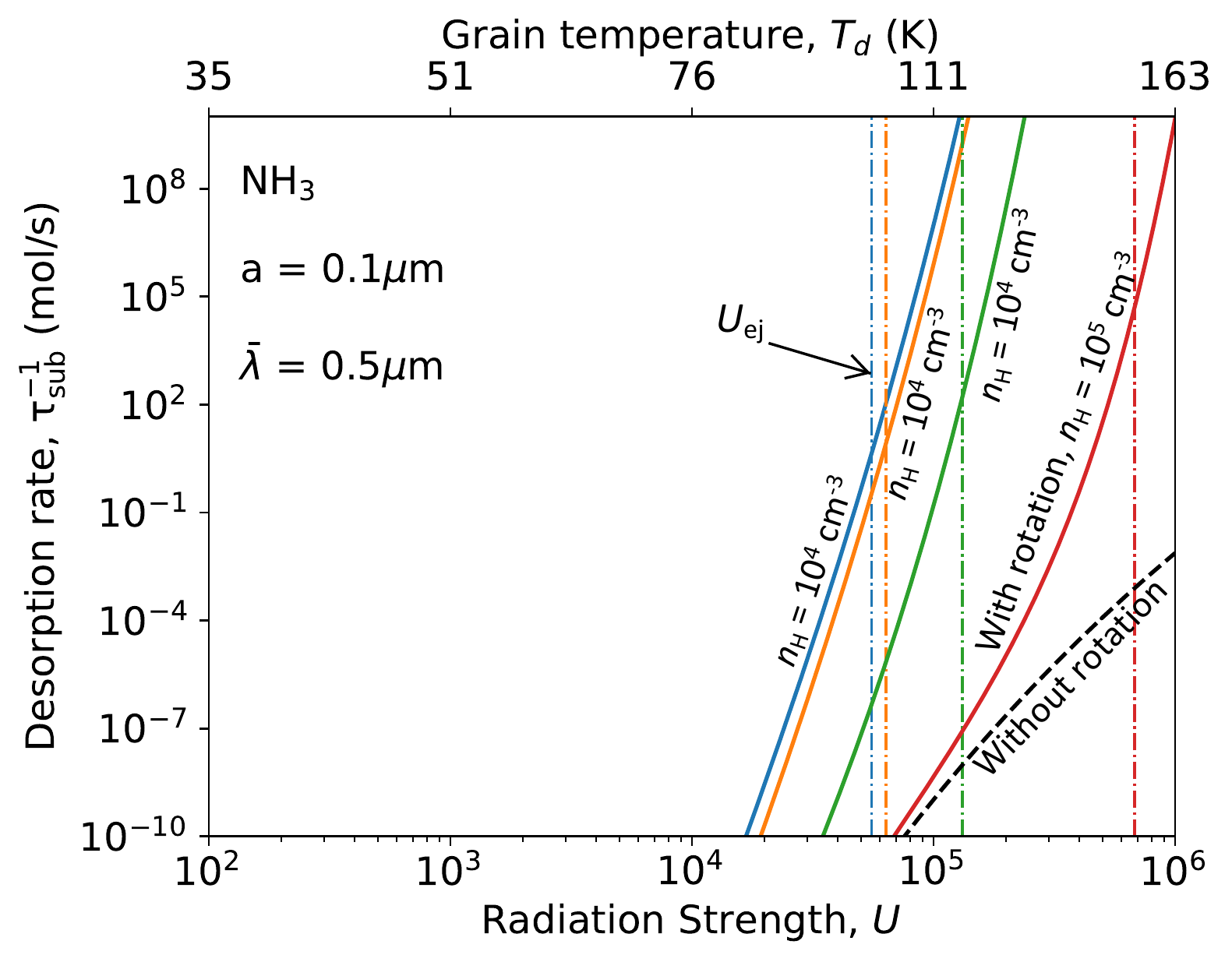}
\caption{Same as Figure \ref{fig:rate_U_wave05_a02} but for N$_{2}$ and NH$_{3}$ molecules, assuming $a=0.2\mum$ (upper panels) and $a=0.1\mum$ (lower panel). Ro-thermal desorption is much faster than thermal desorption for both molecules.}
\label{fig:rate_U_N}
\end{figure*}

Figure \ref{fig:rate_U_N} shows results for N$_{2}$ and NH$_{3}$ molecules, assuming $a=0.1\mum$ and $0.2\mum$. The similar trend as other molecules (Figure \ref{fig:rate_U_wave05_a02}) is observed. The efficiency of ro-thermal desorption is clearly seen for NH$_3$ which has high sublimation temperature.

\begin{figure*}
\includegraphics[width=0.5\textwidth]{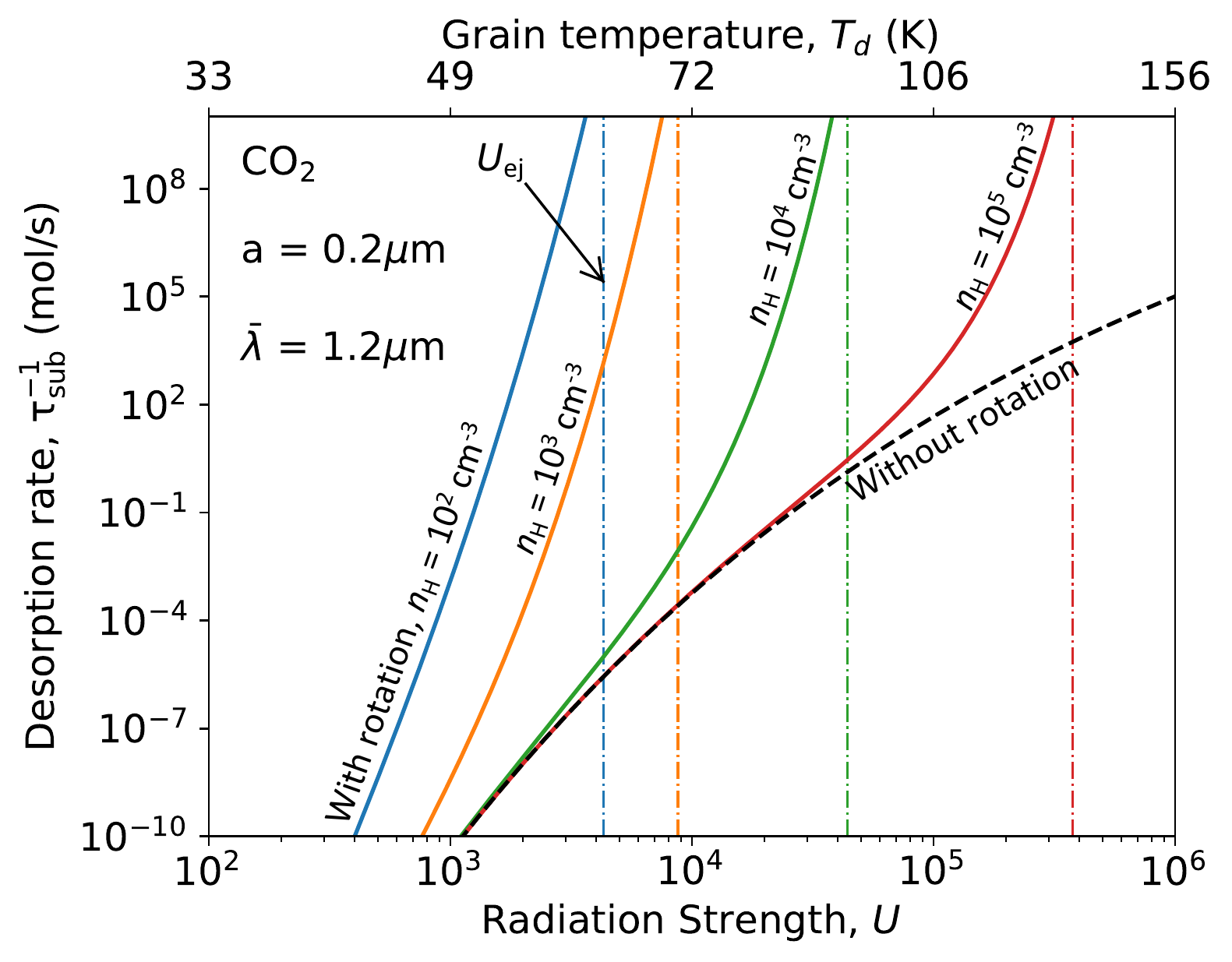}
\includegraphics[width=0.5\textwidth]{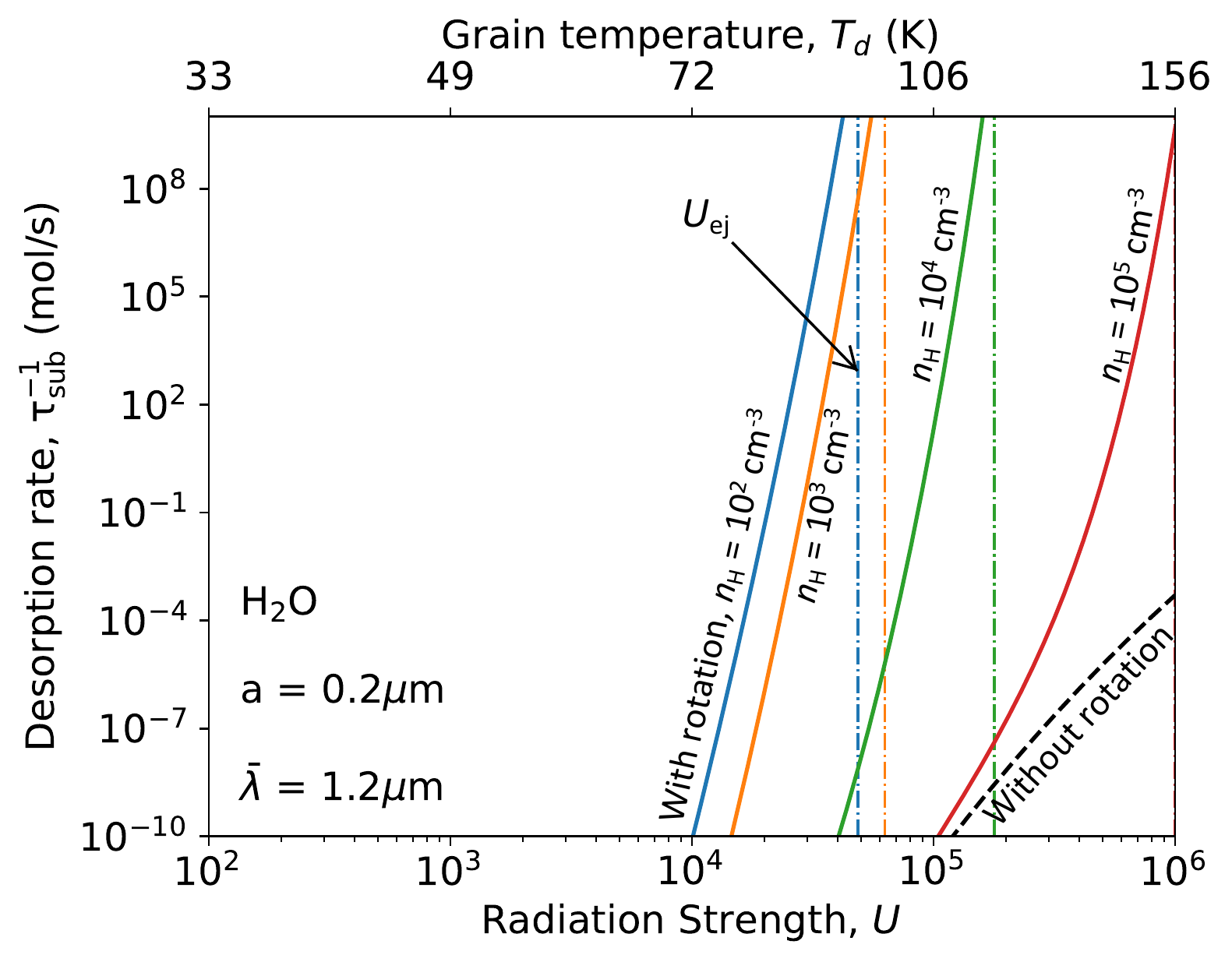}
\includegraphics[width=0.5\textwidth]{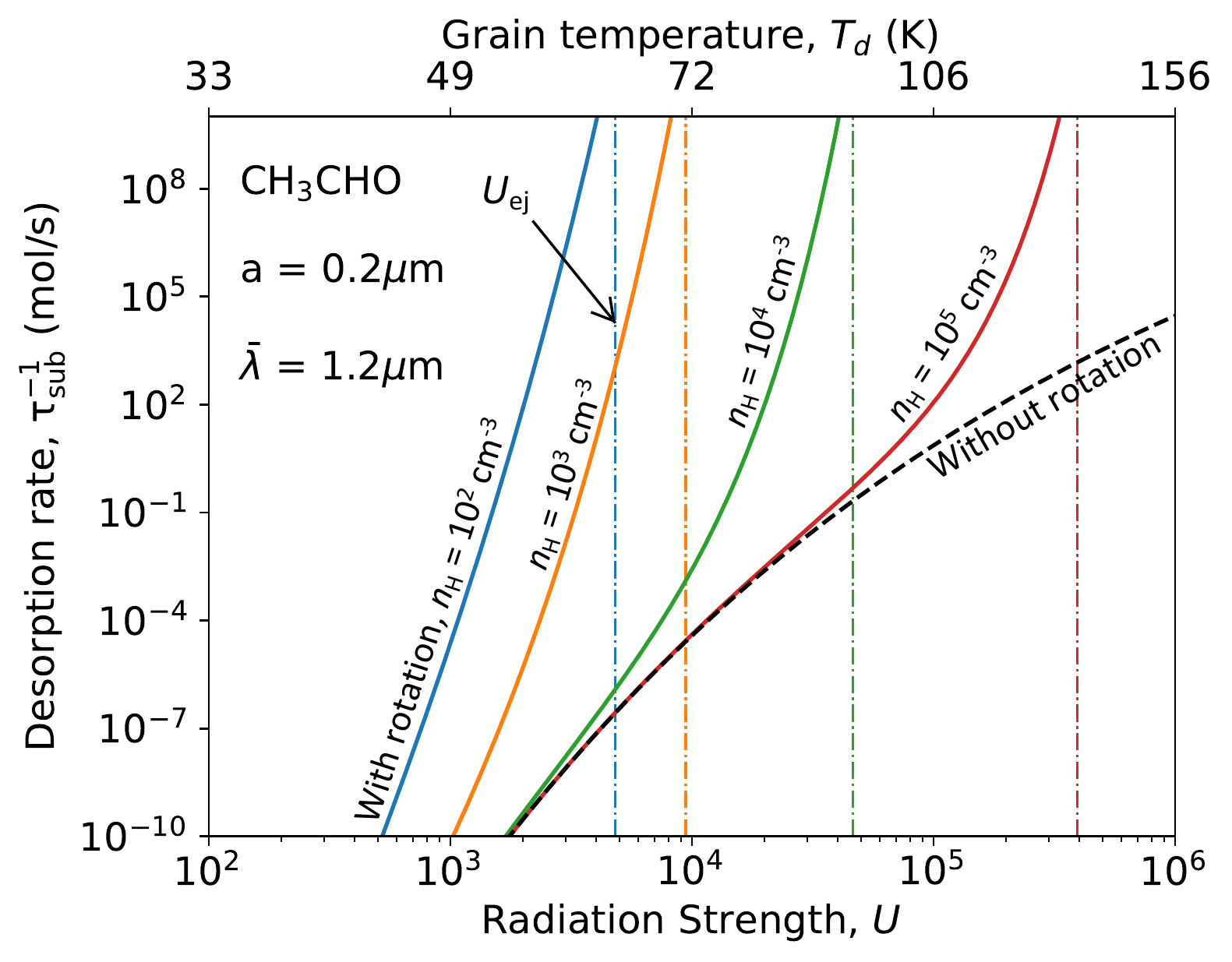}
\includegraphics[width=0.5\textwidth]{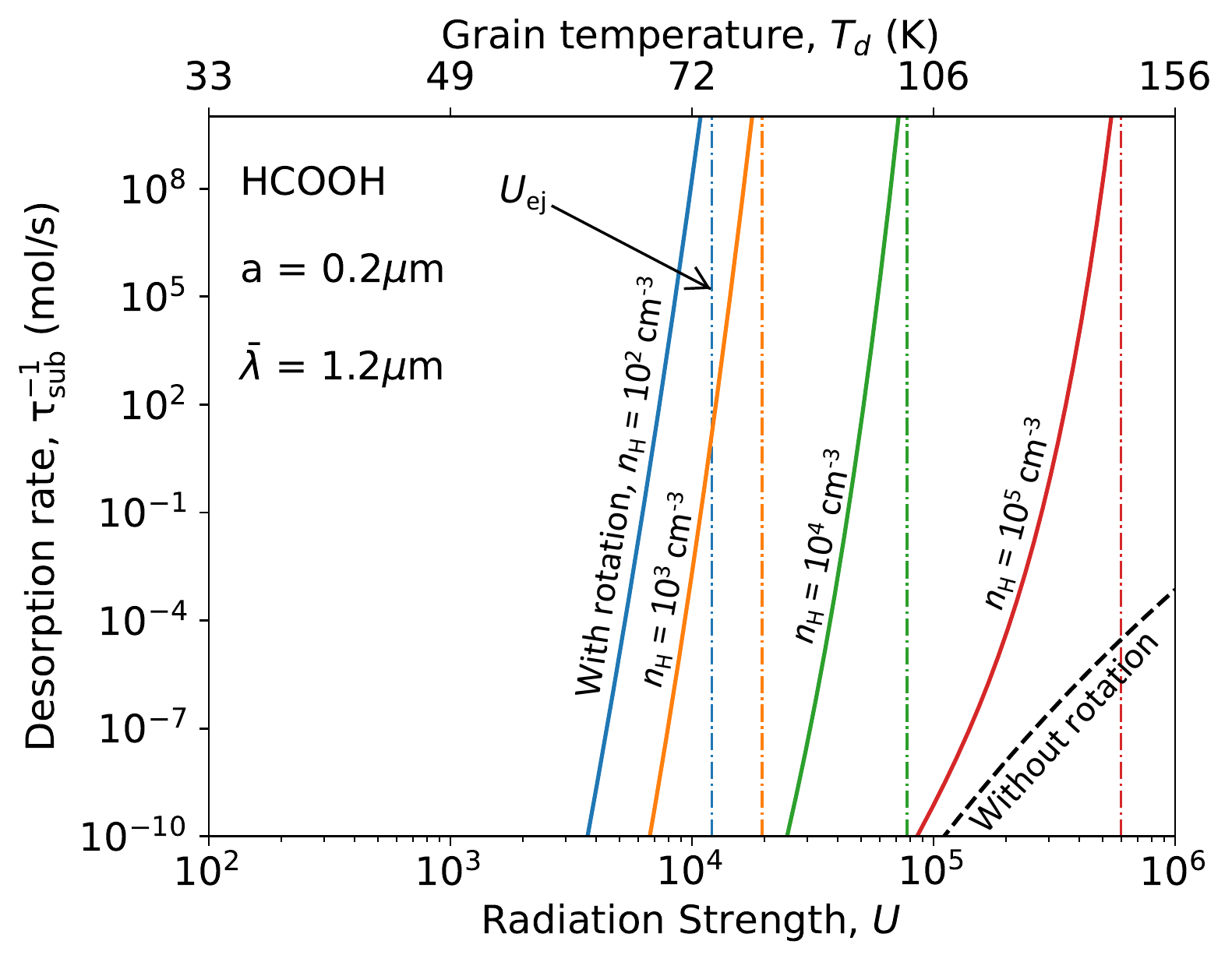}
\includegraphics[width=0.5\textwidth]{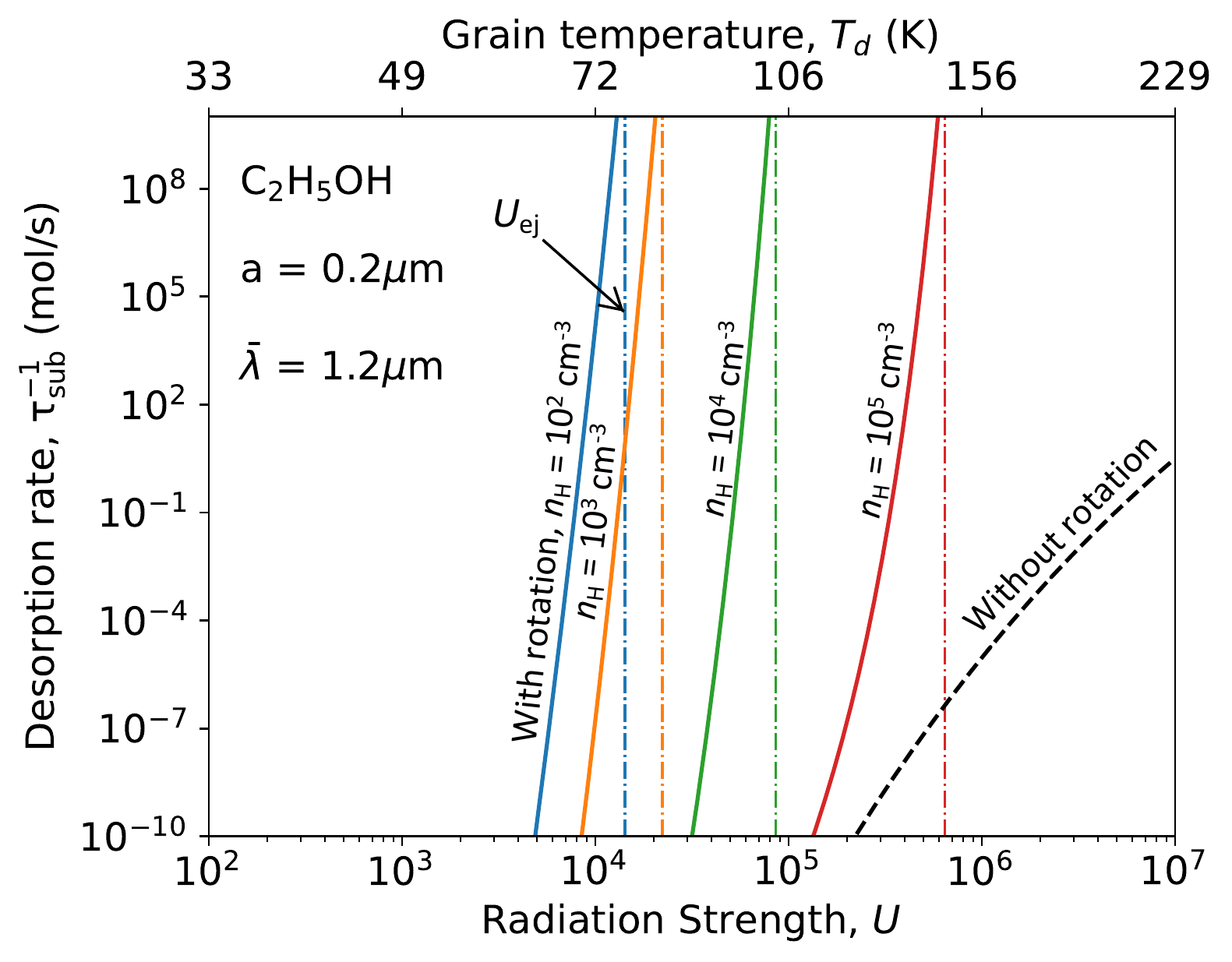}
\includegraphics[width=0.5\textwidth]{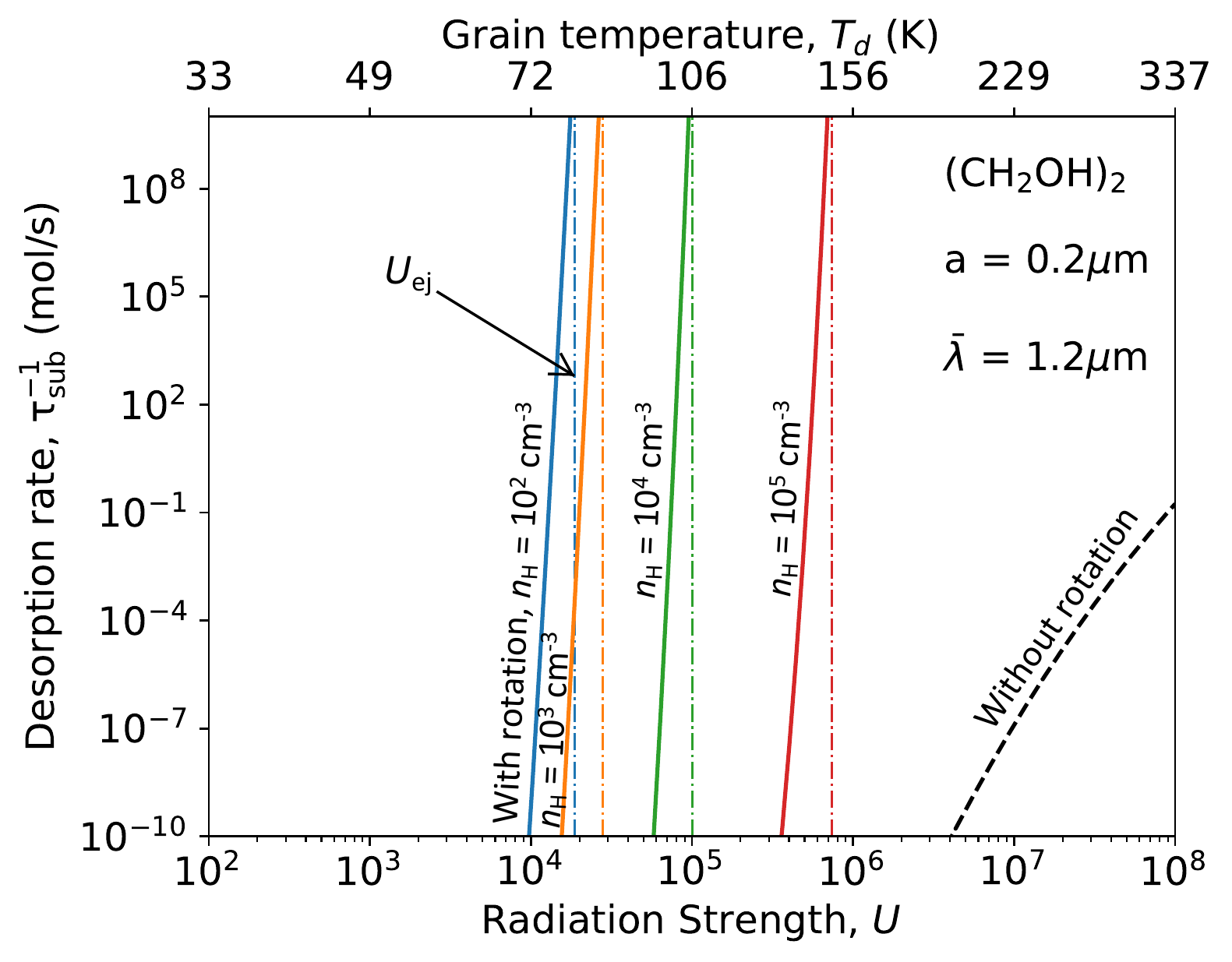}
\caption{Same as Figure \ref{fig:rate_U_wave05_a02} but for the mean wavelength $\bar{\lambda}=1.2\mum$. Ro-thermal desorption appears to be much faster than thermal desorption.}
\label{fig:rate_U_wave12_a02}
\end{figure*}

Same as Figure \ref{fig:rate_U_wave05_a02}, but Figure \ref{fig:rate_U_wave12_a02} shows the results for attenuated radiation fields with $\bar{\lambda}=1.2\mum$. The efficiency of ro-thermal desorption is weaker than the case of $\bar{\lambda}=0.5\mum$, but still dominates over thermal desorption.

\begin{figure*}
\includegraphics[width=0.5\textwidth]{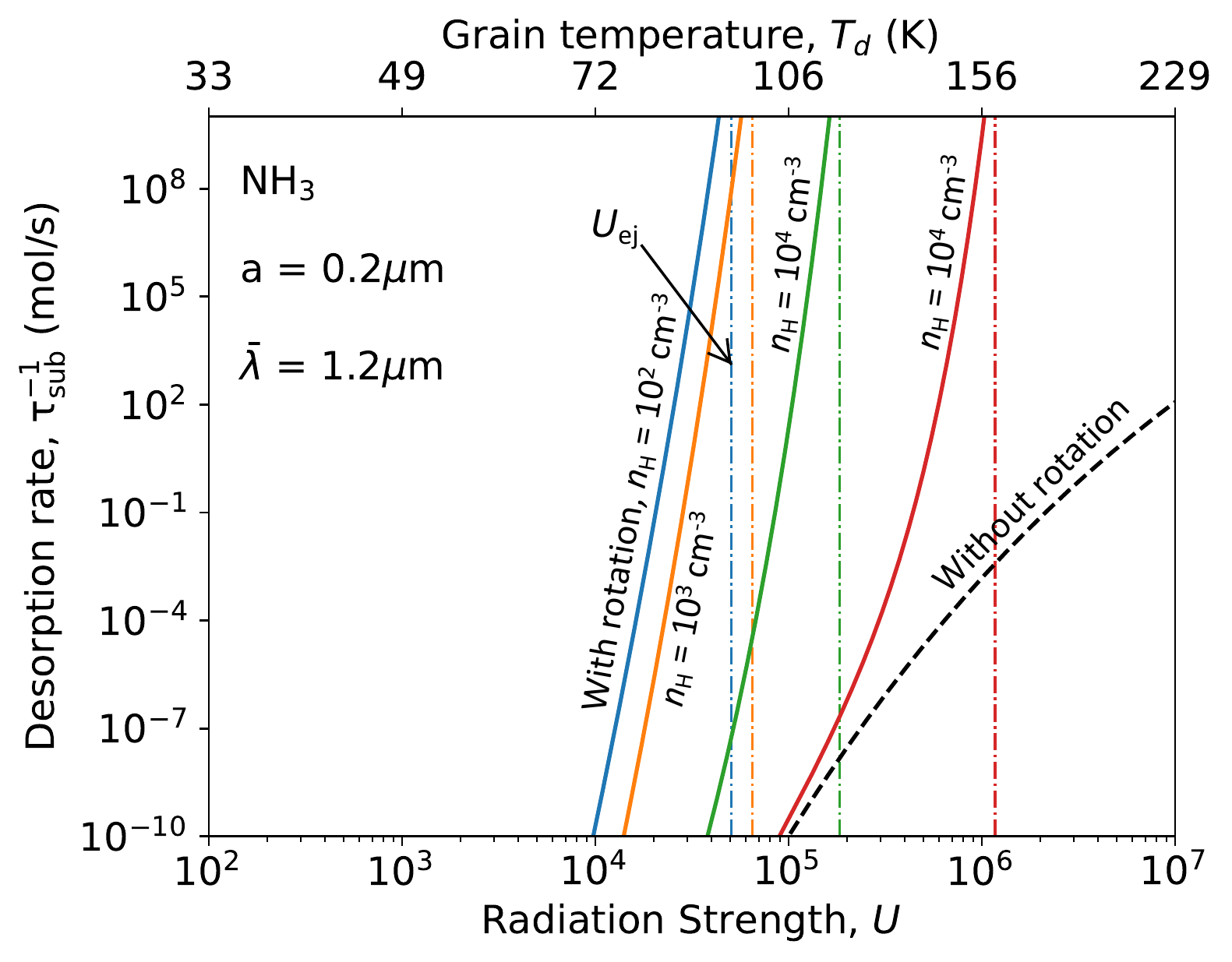}
\includegraphics[width=0.5\textwidth]{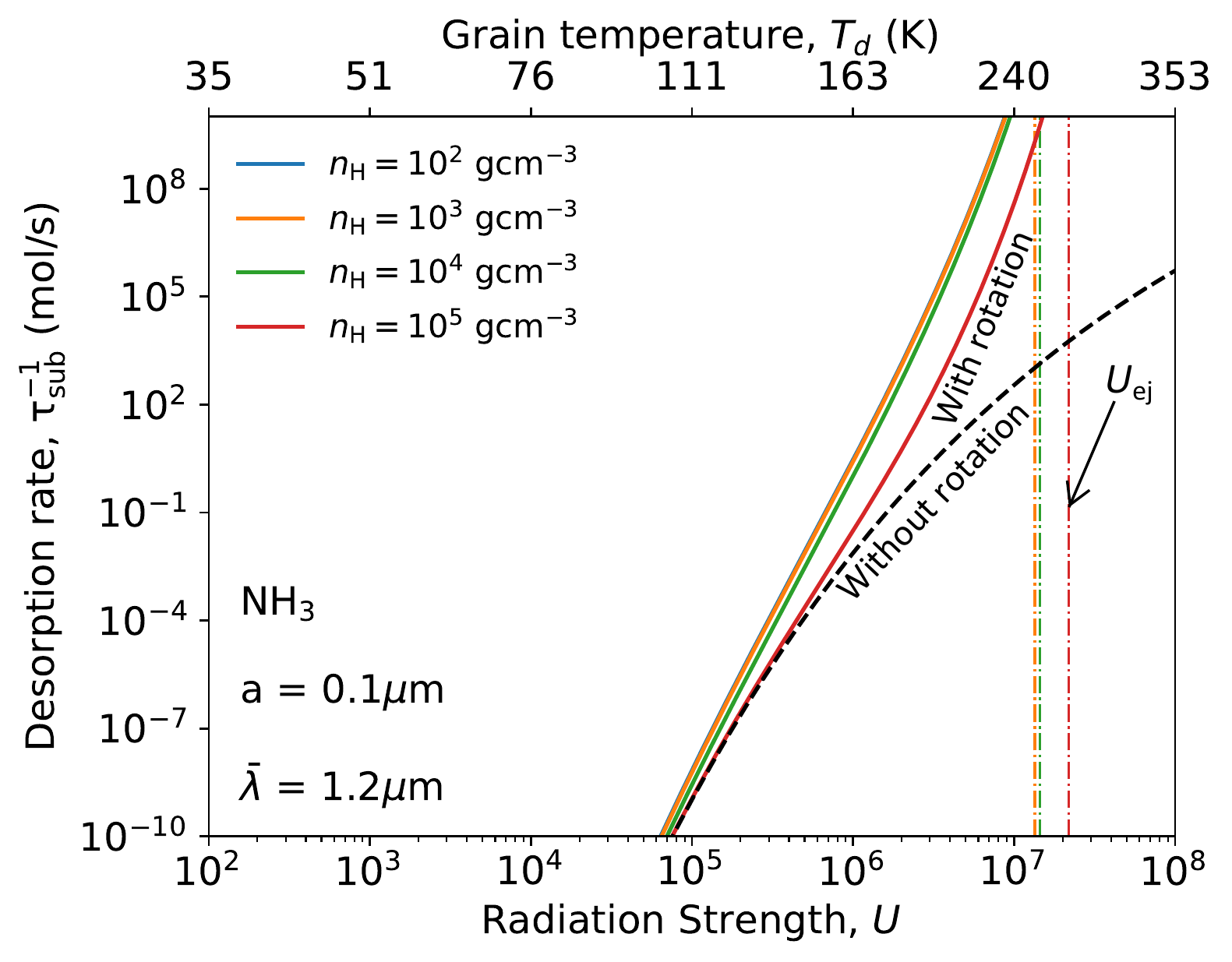}
\includegraphics[width=0.5\textwidth]{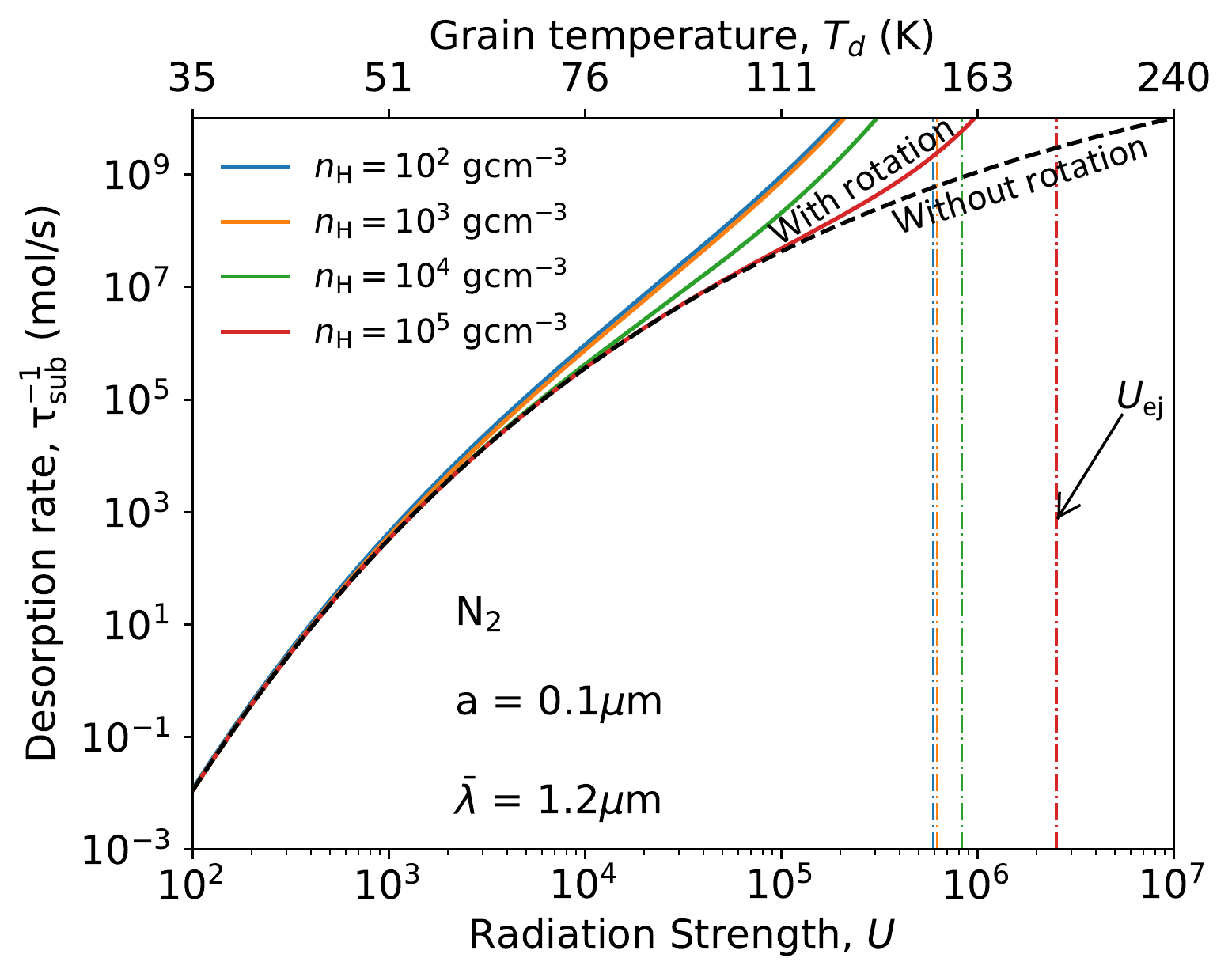}
\includegraphics[width=0.5\textwidth]{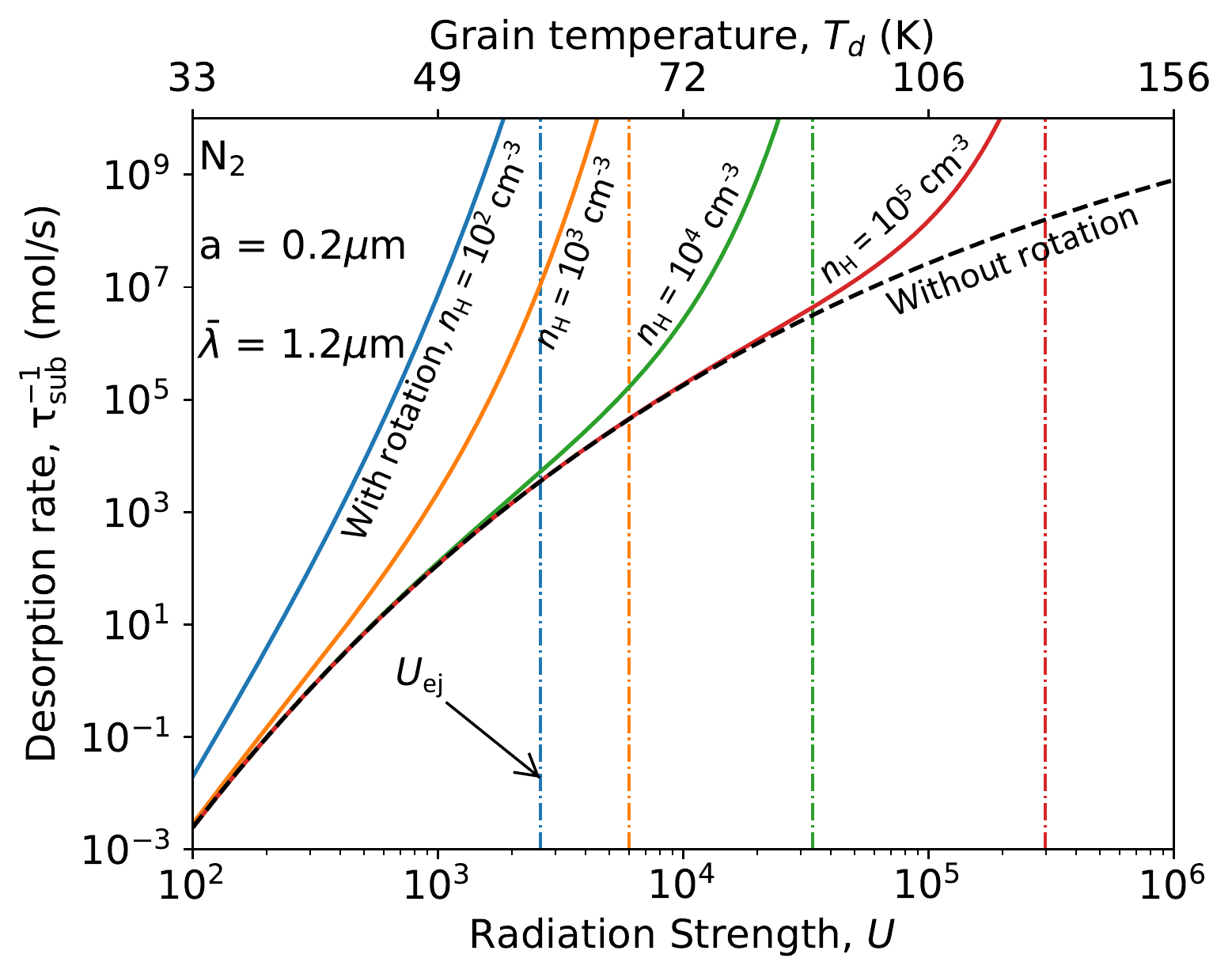}
\caption{Same as Figure \ref{fig:rate_U_wave12_a02} but for N$_{2}$ and NH$_{3}$ molecules. Ro-thermal desorption is much faster than thermal desorption for both molecules.}
\label{fig:rate_U_N2}
\end{figure*}

Figure \ref{fig:rate_U_N2} shows similar results as Figure \ref{fig:rate_U_wave12_a02} but for $\bar{\lambda}=1.2\mum$. The results are slightly different due to the radiation field with longer mean wavelength $\bar{\lambda}$.

\subsubsection{Temperature threshold for ro-thermal vs. thermal desorption}
\begin{figure*}
\includegraphics[width=0.5\textwidth]{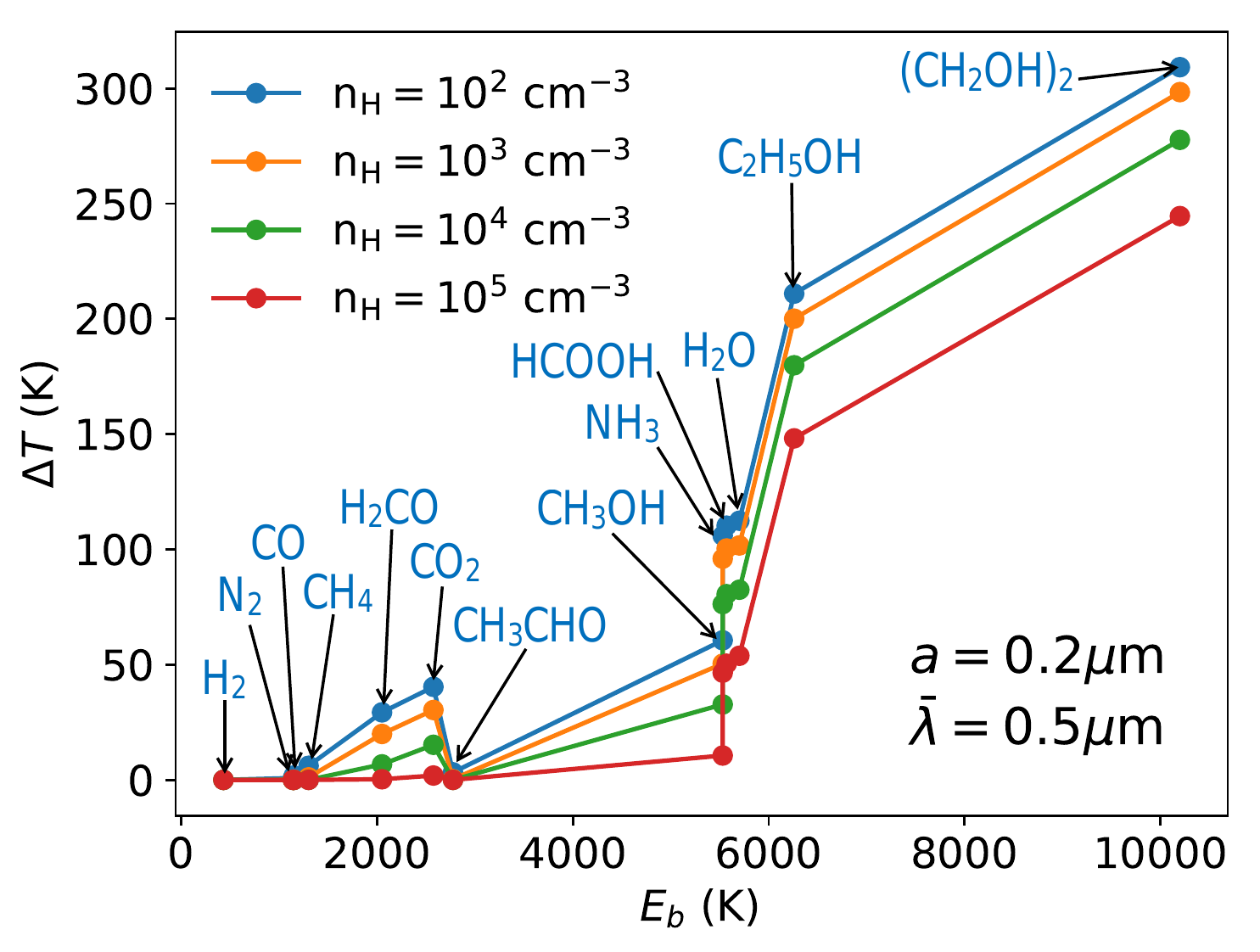}
\includegraphics[width=0.5\textwidth]{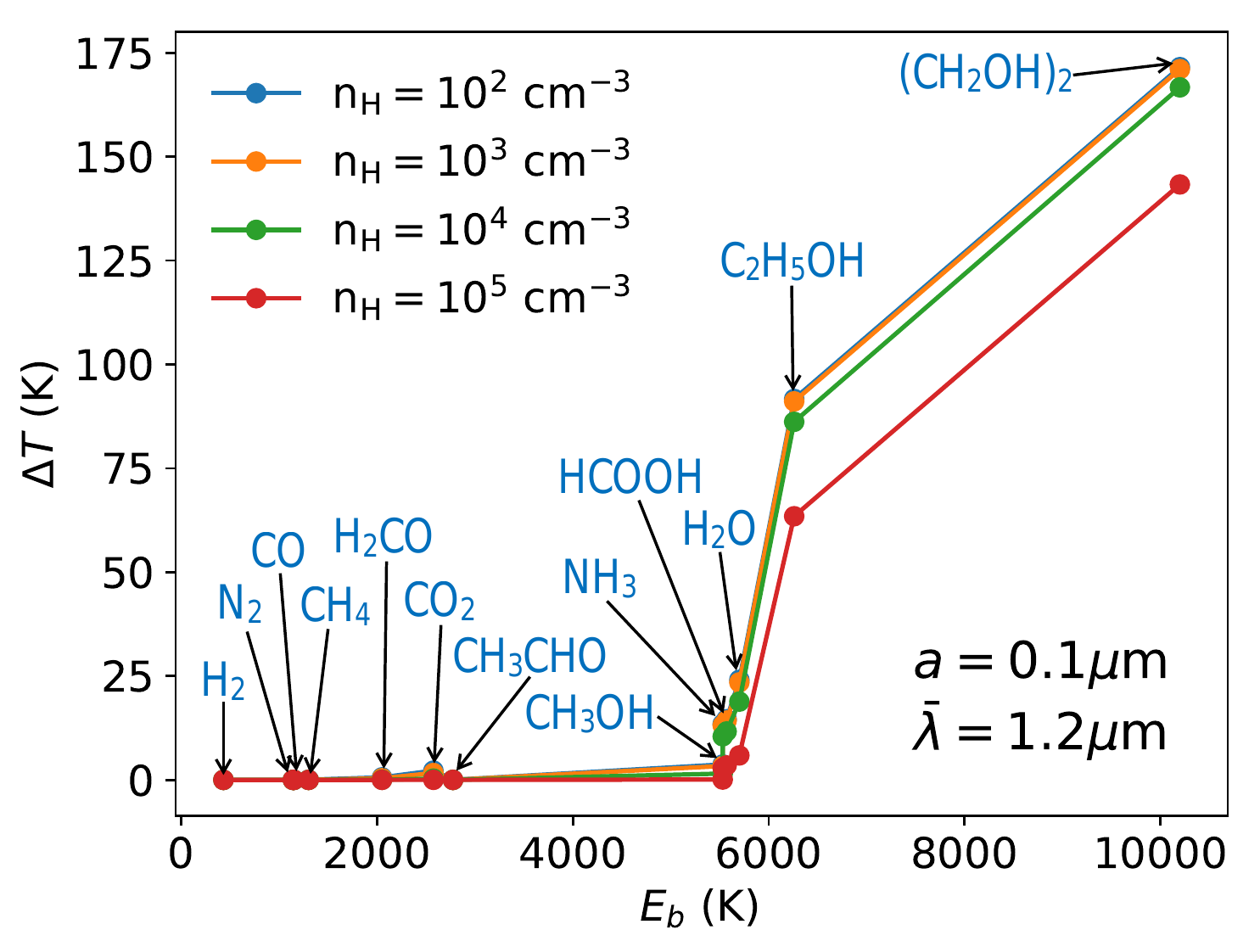}
\includegraphics[width=0.5\textwidth]{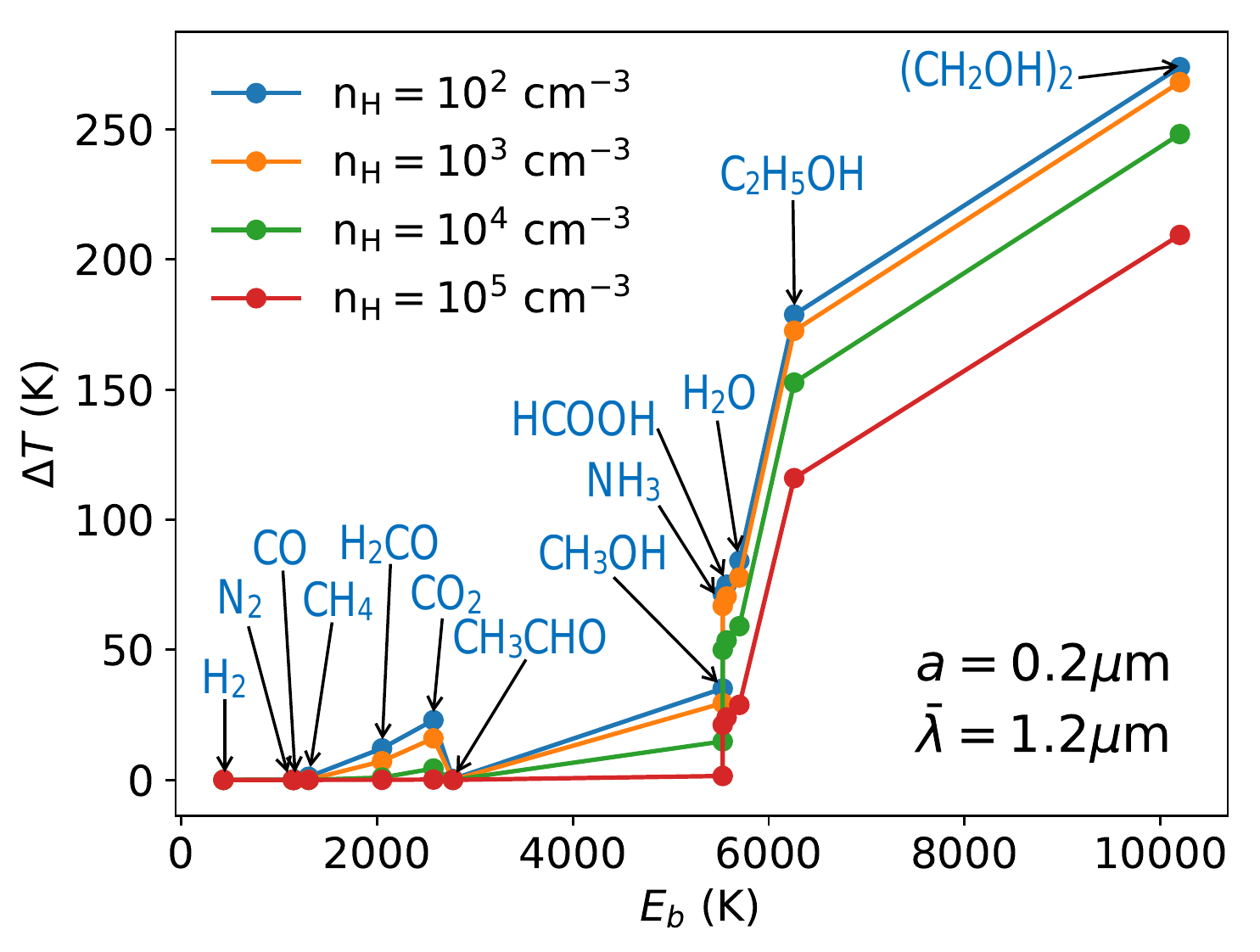}
\includegraphics[width=0.5\textwidth]{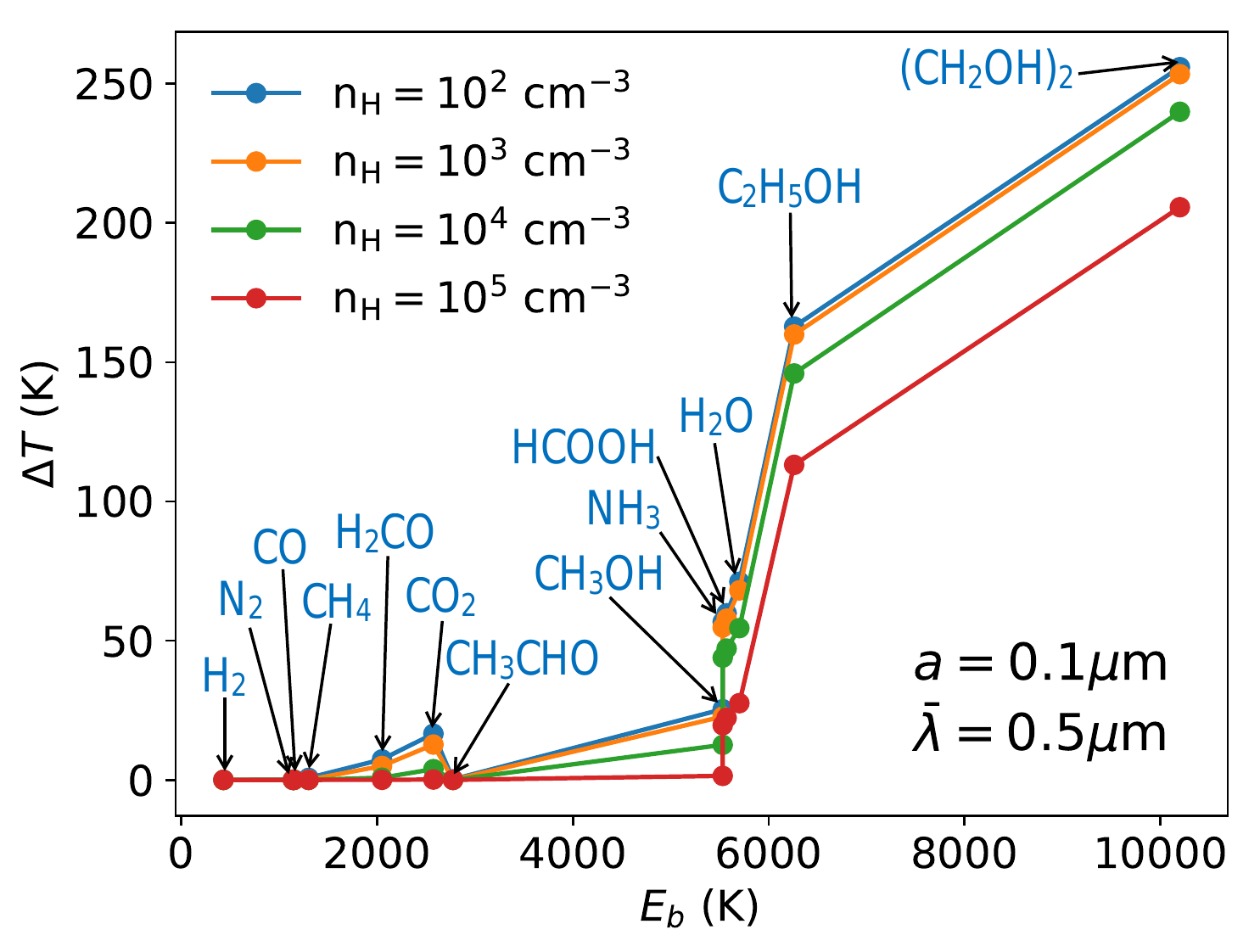}

\caption{Difference between the temperature of ro-thermal desorption and that of thermal sublimation, $\Delta T=|T_{\rm sub,rot}-T_{\rm sub,0}$, as a function of molecule binding energy for $a=0.1\mum$ and $a=0.2\mum$ for $\bar{\lambda}=0.5\mum$ and $1.2\mum$. The difference increases with increasing $E_{b}$ and with the gas density.}
\label{fig:deltaT}
\end{figure*}

Figure \ref{fig:deltaT} shows the decrease of sublimation temperature, $\Delta T=|T_{\rm sub,rot}-T_{\rm sub,0}|$ due to centrifugal potential as a function of the radiation intensity ($U$) and the grain temperature, assuming the different gas density. Analytical results from Equation (\ref{eq:dTsub}) are also shown for comparison. The effect of ro-thermal desorption is more important for lower density. Ro-thermal desorption is also more efficient for molecules with higher binding energy where ro-thermal desorption can occur at more than 100 K lower than the thermal desorption. The efficiency of ro-thermal desorption is more efficient for stellar photons of $\bar{\lambda}=0.5\mum$ but less efficient for reddened photons with $\bar{\lambda}=1.2\mum$.

At high densities of $n_{\rm H}=10^{5}\cm^{-3}$, ro-thermal desorption can still occur at temperatures much lower than thermal desorption for water and other molecules with high binding energy.

Note that the molecule CO$_2$ has low $E_{b}\sim 2575\K$ but its sublimation temperature is high of $T_{\rm sub,0}\sim 72\K$ (\citealt{2004MNRAS.354.1133C}), which results in a slightly peaky feature $\Delta T$ in Figure \ref{fig:deltaT}.

\section{Discussion}\label{sec:discuss}
\subsection{Thermal desorption, rotational desorption, ro-thermal desorption}
Thermal desorption (sublimation) is a popular mechanism to release water and complex organic molecules from the icy grain mantle in star-forming regions \citep{Herbst:2009go}. This desorption process ignores the fact that grains are rapidly spinning due to radiative torques when grains are subject to intense radiation field from protostars (\citealt{1996ApJ...470..551D}; \citealt{Hoang:2008gb}; \citealt{2009ApJ...695.1457H}; \citealt{Herranen:2019kj}). 

The effect of suprathermal rotation on the desorption of molecules from the grain surface is first studied in \cite{Hoang:2019td} where the authors discovered that the resulting centrifugal force is sufficient to disrupt the ice mantle into small fragments. Subsequently, molecules can evaporate from these fragments due to transient heating by UV photons or enhanced thermal sublimation. This {\it rotational} desorption mechanism is found to be efficient in hot cores/corinos around young stars where the radiation strength $U$ can reach $U\sim 10^{8}-10^{9}$ (e.g., grain temperature $T\sim 500\K$). The efficiency of rotational desorption increases with the thickness of the ice mantle, so it is most efficient for grains with a thick ice mantle.

In this paper, we study the additional effect of grain rotation on thermal desorption in regions where grain temperatures are below the sublimation threshold of water and molecules, which have $U<10^{6}$ or $T<200\K$. Specifically grain rotation provides molecules on the grain surface with a centrifugal force that acts in the opposite direction from the binding force. As a result, a rather weak level of thermal excitation can help molecules to sublimate if grains are spinning rapidly. We term this mechanism {\it ro-thermal desorption} mechanism. {The difference between ro-thermal desorption and rotational desorption is that in the former process molecules sublimate directly from the intact icy grain mantle, whereas in the latter process, the ice mantle is first disrupted into tiny icy fragments and subsequently molecules sublimate from these icy fragments.} 

The efficiency of the ro-thermal desorption depends both on the grain rotation rate and grain temperature, but the rotation rate plays a key role. Therefore, ro-thermal desorption can occur in weak radiation fields when grains can be spun-up by mechanical torques (\citealt{2007ApJ...669L..77L}; \citealt{2018ApJ...852..129H}). The ro-thermal desorption mechanism takes over rotational desorption when the ice mantle is thin (i.e., $\Delta a_{m}\ll a$) such that the tensile stress acting on the ice mantle is rather small and cannot desorb the entire mantle (see Figures \ref{fig:rate_U_wave05_a02}-\ref{fig:rate_U_wave05_a01}). 

We also find that individual molecules can be directly ejected from the icy grain mantle for $\omega\gtrsim \omega_{\rm ej}$, and this rotational desorption process requires higher radiation strength than ro-thermal desorption (see Figures \ref{fig:rate_U_wave05_a02} and \ref{fig:rate_U_wave05_a01}). Compared to rotational desorption, we find that ro-thermal desorption occurs at lower radiation strength before the entire mantle can be disrupted into small fragments and efficient for thin ice mantle of thickness $l<100\AA$. 

Compared to UV photodesorption that requires FUV photons between $7-10.5$ eV to be effective (\citealt{2007ApJ...662L..23O}; \citealt{vanDishoeck:2013en}), ro-thermal desorption can work with optical photons with the mean wavelength even at $\bar{\lambda}\gtrsim 0.5\mum$. Therefore, ro-thermal desorption can be effective in regions without FUV.

{Finally, the rot-thermal desorption mechanism is expected to be most efficient in astrophysical environments with $U/n_{\H}T_{\gas}^{1/2}> 1$ (see Eqs. \ref{eq:omega_RAT1} and \ref{eq:omegaRAT2}), such as star-forming regions, the surface and intermediate layer of protoplanetary disks, reflection nebula (e.g., \citealt{2015MNRAS.448.1178H}), and photodissociation regions.}

\subsection{Ro-thermal desorption of PAHs}
Like other molecules, PAHs condense in the ice mantle of dust grains in cold dense clouds (\citealt{1999Sci...283.1135B}; \citealt{2014A&A...562A..22C}; \citealt{2015ApJ...799...14C}). Yet, the question of how PAHs are returned into the gas phase is still unclear.

Ro-thermal desorption appears to be an efficient mechanism to desorb PAHs. Since ro-thermal desorption requires lower radiation strength to desorb than rotational desorption, one can describe the efficiency of ro-thermal desorption by considering the ejection threshold. Using Equation (\ref{eq:omega_ej}), one obtains the ejection threshold of PAHs:
\bea
\omega_{\rm ej}=\left(\frac{3E_{b}}{ma^{2}}\right)^{1/2}\simeq \frac{10^{10}}{a_{-5}}\left(\frac{(E_{b}/k)}{4000\K}\frac{m_{\rm C6H6}}{m}\right)^{1/2}\rm rad\s^{-1},\label{eq:omega_ej_PAH}~~~~
\ena
where the binding energy of benzene C$_{6}$H$_{6}$ and naphthalene (C$_{10}$H$_{8}$) to ice is $E_{b}/k\sim 4000\K$ and $6000\K$ (see Table 4 and 5 in \citealt{Michoulier:2018cx}).
 
The ejection radiation strength is then
\bea
U_{\rm ej}\simeq 35n_{1}T_{2}^{1/2}(1+F_{\rm IR})\frac{\lambda_{0.5}^{1.7}}{\gamma a_{-5}^{1.7}}\left(\frac{(E_{b}/k)}{4000\K}\frac{m_{\rm C6H6}}{m}\right)^{1/2}~\label{eq:Uej_PAH}
\ena
for $a\lesssim a_{\rm trans}$, and 
\bea
U_{\rm ej}\simeq 1.8n_{1}T_{2}^{1/2}(1+F_{\rm IR})\frac{\lambda_{0.5}^{1.7} a_{-5}}{\gamma}\left(\frac{(E_{b}/k)}{4000\K}\frac{m_{\rm C6H6}}{m}\right)^{1/2}
\ena
for $a> a_{\rm trans}$. 
Clearly, the ejection threshold is much lower than that of water and COMs (see Figures \ref{fig:rate_U_wave05_a02}-\ref{fig:rate_U_wave12_a02}). Therefore, the ro-thermal desorption is efficient for desorption of PAHs in star-forming regions.

\subsection{Ro-thermal desorption in photodissociation regions and protoplanetary disks}
Photodissociation regions (PDRs) are traditionally dense molecular clouds with typical gas density $n_{\H}\lesssim 10^{5}\cm^{-3}$ illuminated by O or B stars (e.g., Orion Bar) with radiation strength $U\lesssim 10^{5}-10^{6}$ (see \citealt{1999RvMP...71..173H}; \citealt{Tielens:2007wo}). 

For a typical PDR model like Orion Bar (\citealt{2005ApJ...630..368A}), the gas density and radiation fields are $n\sim 10^{4}\cm^{-3}$ and  $U\sim 3\times 10^{4}$. For these physical conditions, from Figures \ref{fig:rate_U_wave05_a02} and \ref{fig:rate_U_wave05_a01}, we see that ro-thermal desorption is very efficient in desorbing water and complex organic molecules. This mechanism can explain the formation of COMs and PAHs which are usually observed in PDRs (see \citealt{2017SSRv..212....1C} for a review).

The surface and intermediate layer of protoplanetary disks around young stars are attractive targets for studying ro-thermal desorption due to strong radiation fields. For the same radiation intensity, the rate of ro-thermal desorption is several orders of magnitude higher than that of thermal desorption (\citealt{1995ApJ...455L.167S}).

We note that in more intense radiation field of hot cores/corinos, the entire ice mantle could be desorbed via {\it rotational desorption} mechanism \citep{Hoang:2019td}. {Finally, with this paper, the effect of grain rotation on thermal desorption of molecules is complete and for the first time demonstrate the importance of accounting for grain dynamics for grain-surface chemistry.}

{We note that grains can also be spun-up by mechanical torques (\citealt{2007ApJ...669L..77L}; \citealt{2018ApJ...852..129H}). Therefore, ro-thermal desorption can occur for supersonic flows. The potential environments include icy grains in young stellar outflows (see \citealt{Hoang:2019td}). 
}

\subsection{From experimental data to astrochemical modeling}
Astrochemical modeling of observational data usually takes desorption rates and chemical reaction rates measured from experiment and apply directly to interstellar dust grains (e.g., \citealt{Oberg:2009dp}). Usually, the physical properties of dust grains, including grain surface, grain size, and grain dynamics, are disregarded (see \citealt{Caselli:2012fq}). Recently, the effect of grain surface properties was studied by experiments in \cite{Potapov:2019wg}, but the application to the specific grain surface is not yet available.
 
In light of our findings, application of experimental measurements of sublimation temperatures cannot be directly applied to model thermal sublimation of molecules from the grain mantle due to the effect of grain suprathermal rotation. We find that the effective sublimation temperature is much lower than the measured temperature for non-rotating grains in the lab, which depends on the local gas density. The difference is significant for water ice and COMs with high binding energy.

\section{Summary}\label{sec:sum}
We have studied the effect of grain suprathermal rotation on the thermal desorption of molecules from icy grain mantles. Our results are summarized as follows:
\begin{itemize}
\item[1]
We find that the centrifugal potential energy due to grain rotation acts to reduce the potential barrier of the molecule desorption and formulate a theory for thermal desorption of molecules from rapidly spinning grains. To differentiate from the classical thermal desorption mechanism, we term this mechanism rotational-thermal desorption or ro-thermal desorption.

\item[2] We apply the ro-thermal desorption theory for icy grains spun-up by radiative torques and find that the rate of ro-thermal desorption of water and COMs is much larger than that of the classical thermal sublimation.

\item[3] We derive the effective temperature threshold for ro-thermal desorption and find that this temperature is much lower than that for thermal desorption. The ro-thermal desorption temperatures decreases with increasing the radiation strength and with decreasing the gas density.

\item[4] We find that COMs can be released via ro-thermal desorption in environments with low temperatures ($T<100 \K$) provided that the gas density is not very high, i.e., $n_{\rm H}< 10^{8}\cm^{-3}$. As a result, interpretation of the detection of COMs in astrophysical conditions by means of grain heating only is likely inadequate because of centrifugal force effect.

\item[5] {To use molecules as a reliable tracer of physical and chemical properties of astrophysical environments,} one needs to take into account the effect of suprathermal rotation of icy grains on desorption of molecules, and chemical modeling should take into account this effect.

\item[6] Our results reveal that using experimental data for astrochemical modeling of gas-grain surface chemistry in star-forming regions would be cautious and must account for the effect of suprathermal rotation of icy grains.

\end{itemize}
\acknowledgments
{We are grateful to the anonymous referees for helpful comments that improved the presentation of the manuscript.} We thank Le Ngoc Tram for useful comments. This work was supported by the National Research Foundation of Korea (NRF) grants funded by the Korea government (MSIT) (2017R1D1A1B03035359 and 2019R1A2C1087045).


\bibliography{ms.bbl}
\end{document}